\documentclass[iop]{emulateapj}
\usepackage{apjfonts,ulem}
\usepackage[backref,breaklinks,colorlinks,citecolor=blue,linkcolor=magenta,urlcolor=blue]{hyperref}

\shorttitle{Cyclotron Resonance in Solar Wind Turbulence}
\shortauthors{Woodham et al.}

\begin{document}

\title{The Role of Proton-Cyclotron Resonance as a Dissipation Mechanism in Solar Wind Turbulence: A Statistical Study at Ion-Kinetic Scales}

\author{Lloyd D. Woodham}
\affiliation{Mullard Space Science Laboratory, University College London, Holmbury St. Mary, Dorking, Surrey, RH5 6NT, UK}
\author{Robert T. Wicks}
\affiliation{Mullard Space Science Laboratory, University College London, Holmbury St. Mary, Dorking, Surrey, RH5 6NT, UK}
\affiliation{Institute for Risk and Disaster Reduction, University College London, Gower Steet, London, WC1E 6BT, UK}
\author{Daniel Verscharen}
\affiliation{Mullard Space Science Laboratory, University College London, Holmbury St. Mary, Dorking, Surrey, RH5 6NT, UK}
\affiliation{Space Science Center, University of New Hampshire, Durham, NH 03824, USA}
\author{Christopher J. Owen}
\affiliation{Mullard Space Science Laboratory, University College London, Holmbury St. Mary, Dorking, Surrey, RH5 6NT, UK}

\email{lloyd.woodham.16@ucl.ac.uk, woodhamlloyd@gmail.com}

\begin{abstract}

We use magnetic field and ion moment data from the MFI and SWE instruments onboard the \textit{Wind} spacecraft to study the nature of solar wind turbulence at ion-kinetic scales. We analyze the spectral properties of magnetic field fluctuations between 0.1 and 5.5 Hz over 2012 using an automated routine, computing high-resolution 92 s power and magnetic helicity spectra. To ensure the spectral features are physical, we make the first in-flight measurement of the MFI `noise-floor' using tail-lobe crossings of the Earth's magnetosphere during early 2004. We utilize Taylor's hypothesis to Doppler-shift into the spacecraft frequency frame, finding that the spectral break observed at these frequencies is best associated with the proton-cyclotron resonance scale, $1/k_c$, compared to the proton inertial length $d_i$ and proton gyroscale $\rho_i$. This agreement is strongest when we consider periods where $\beta_{i,\perp}\sim1$, and is consistent with a spectral break at $d_i$ for $\beta_{i,\perp}\ll1$ and $\rho_i$ for $\beta_{i,\perp}\gg1$. We also find that the coherent magnetic helicity signature observed at these frequencies is bounded at low frequencies by $1/k_c$ and its absolute value reaches a maximum at $\rho_i$. These results hold in both slow and fast wind streams, but with a better correlation in the more Alfv\'enic fast wind where the helicity signature is strongest. We conclude that these findings are consistent with proton-cyclotron resonance as an important mechanism for dissipation of turbulent energy in the solar wind, occurring at least half the time in our selected interval. However, we do not rule out additional mechanisms.

\end{abstract}


\section{Introduction} \label{sec:intro}

The solar wind supports a turbulent energy cascade where the spectrum of magnetic field fluctuations follows a Kolmogorov inertial range scaling of $k^{-5/3}$ extending over several decades \citep{Tu1995,Goldstein1995,Bruno2013}. We convert the wavenumber, $k$, of the turbulent fluctuations along the sampling direction of the solar wind flow to a frequency, $f$, using Taylor's hypothesis \citep{Taylor1938}: $f\sim kv_{sw}/2\pi$, where $v_{sw}$ is the solar wind speed. At frequencies in the plasma frame of order the ion gyrofrequency, $\Omega_i=q_iB_0/m_i$, typically measured around 0.1-1 Hz in the spacecraft frame at 1 AU, the spectrum steepens \citep[e.g.,][]{ColemanJr.1968,Russell1972}. Here, $q_i$ is the ion charge, $m_i$ is the ion mass, and $B_0$ is the background field strength. The observed spectral break in the magnetic field power spectra at these so-called ion-kinetic frequencies has been attributed to the onset of kinetic effects such as dispersion or turbulent dissipation \citep[see][and references therein]{Alexandrova2013,Kiyani2015,Chen2016}, although the actual physical mechanisms behind the steepening remain poorly understood.

In-situ data from spacecraft have revealed a bimodal distribution in solar wind speed with two distinct peaks, leading to the designation of two types of wind: slow ($\sim350 \text{ km s}^{-1}$) and fast ($\sim600 \text{ km s}^{-1}$), attributed to different source regions in the solar corona \citep[e.g.,][]{Schwenn1990,Habbal1997OriginsWind}. In fast wind streams, the spectral steepening is sometimes associated with the start of a variable transition range spanning less than a decade in frequency \citep[see also the power spectra in \citealt{Kiyani2009GlobalTurbulence,Chen2010}]{Sahraoui2010,Smith2012,Kiyani2013EnhancedTurbulence,Bruno2014a,Bruno2017SolarScales}. The spectral index of the spectrum in this range typically lies between -2 and -4 \citep{Smith2006,Hamilton2008,Koval2013,Bruno2014a}. At even higher frequencies, the spectrum changes again to a second, more `universal' power law of $f^{\;-2.8}$ continuing towards electron scales, associated with dispersive modes such as kinetic Alfv\'en waves (hereafter, KAWs) and whistler waves \citep[e.g.,][]{Gary2009,TenBarge2012InterpretingWind,Boldyrev2013}, or small-scale coherent structures such as current sheets \citep[e.g.,][]{Perri2012DetectionTurbulence}. The slow wind in general, typically lacks a transition range and instead shows a single steepening from $f^{-5/3}$ to about $f^{-2.8}$ \citep{Bruno2014a,Bruno2017SolarScales}.

There is strong evidence of the coupling between magnetic energy in the turbulent fluctuations and kinetic energy of the ions, linking the large-scale turbulent cascade with heating of the solar wind particle distributions. For example, the temperature of the solar wind decreases with radial distance more slowly than expected for adiabatic expansion \citep{Marsch1982b,Richardson1995}, implying an active heating process during its expansion \citep[e.g.,][]{Cranmer2009}, which is consistent with the energy cascade rate throughout the inertial range \citep[e.g.,][]{MacBride2008,Stawarz2009}. In the fast wind, the temperature anisotropy, $T_\perp/T_\parallel$, of the proton core population and plateau formation in the proton velocity distributions \citep{Tu2001OnCorona,Marsch2001,Tu2002,Marsch2004,Heuer2007DiffusionProtons} also suggests ongoing heating by the turbulent cascade. These observations indicate that dissipation of the turbulent fluctuations is a likely candidate for the spectral steepening. In fact, the steepness of spectra is correlated with the energy cascade rate and power level in the inertial range \citep{Smith2006,Bruno2014a}, as well as the thermal proton temperature \citep{Leamon1998a}, implying that steeper slopes are associated with greater heating rates.

The kinetic features of the proton velocity distributions highlight a deviation from local thermal equilibrium that is due to the lack of Coulomb collisions in the solar wind \citep[][see also the review by \citealt{Marsch2006}]{Marsch2012}. Instead, these features are likely regulated by linear and non-linear wave-particle interactions \citep[e.g.,][]{Howes2008,Schekochihin2009,Chandran2010,Smith2012,Osman2014MagneticWind} such as ion-cyclotron resonance, Landau resonance and transit-time damping, stochastic heating, entropy cascades, and reconnection-associated mechanisms. There is also evidence that plasma instabilities play an important role \citep{Kasper2002a,Hellinger2006,Matteini2007,Bale2009,Maruca2012,Osman2013,Servidio2014}. These physical processes may lead to the dissipation of energy from the turbulence and subsequent heating of ions observed by spacecraft. Understanding these mechanisms in the collisionless solar wind plasma is a major outstanding problem in the field of heliophysics research.

\subsection{Spectral Steepening at High Frequencies}

Several different characteristic ion plasma scales have been suggested to correspond to the observed spectral steepening, and each one is associated with different plasma heating processes. Two scales that are commonly proposed to correspond to the spectral break are the ion inertial length, $d_i=v_A/\Omega_i$, and the ion gyroscale, $\rho_i=v_{th,\perp}/\Omega_i$. Here, $v_A=B_0/\,\sqrt[]{\mu_0n_im_i}$ is the Alfv\'en speed, $n_i$ is the ion number density, $v_{th,\perp}=\sqrt[]{2k_BT_{i,\perp}/m_i}$ is the ion thermal speed perpendicular to the background magnetic field, $\mathbf{B}_0$, and $T_{i,\perp}$ is the ion perpendicular temperature. The inertial length is associated with the onset of dispersive effects due to the Hall current term, as well as reconnection of small-scale current sheets \citep{Dmitruk2004,Galtier2006,Galtier2007}, whereas the transition from Alfv\'en wave to KAW-dominated turbulence occurs at scales comparable to the gyroscale \citep{Howes2008,Schekochihin2009,Boldyrev2012}.

Another explanation for the observed spectral steepening is cyclotron resonance of Alfv\'en waves with solar wind ions \citep[e.g.,][]{ColemanJr.1968,Marsch1982,Denskat1983,Goldstein1994,Marsch2003Onwind,Gary2004,Smith2012}. Here, the only ions we consider are protons, and throughout this paper we use the subscript, $i$, to refer exclusively to protons. \citet{Leamon1998a} proposed a wavenumber for the onset of cyclotron damping of Alfv\'en waves \citep[see also][]{Gary1999CollisionlessTheory}. The cyclotron resonance condition for protons is given by equating the Doppler-shifted wave frequency in the plasma frame, $\omega$, to the proton gyrofrequency, $\Omega_i$ \citep[e.g., see][]{Stix1992},

\begin{equation}
\omega(k_\parallel)-k_\parallel v_\parallel=\pm {{\Omega }_{i}},
\end{equation}

\noindent where $v_\parallel$ is the parallel velocity of the resonant protons and $k_{\parallel}$ is the parallel component of the wavenumber with respect to $\mathbf{B}_0$. The $\pm$ sign takes into account the sense of polarization of the wave. The wave electric field vector of left-hand circularly-polarized Alfv\'en/ion-cyclotron waves (hereafter, AICs) propagating parallel to $\mathbf{B}_0$ rotates in the same direction as proton gyration, so we use the positive sign. This interaction is most effective if $k_\parallel v_\parallel<0$, reducing the resonance condition to:

\begin{equation}
\omega(k_\parallel)+k_\parallel \left|v_\parallel\right|={{\Omega }_{i}}.
\end{equation}

\noindent To obtain the minimum wavenumber, $k_\parallel=k_c$, at which dissipation of the waves by cyclotron resonance with the background solar wind proton distribution occurs, we take $v_\parallel= v_{th,\parallel}$, where $v_{th,\parallel}$ is the parallel thermal speed of the proton velocity distribution, and for simplicity, substitute for $\omega(k_\parallel)$ using the wave dispersion-relation of Alfv\'en waves (e.g. \citealt{Gary1993}): $\omega(k_\parallel)=k_{\parallel}v_A$, 

\begin{equation} \label{equ:kc}
{{k}_{c}}=\frac{{{\Omega}_{i}}}{{{v}_{A}}+{{v}_{th,\parallel}}}\equiv\frac{1}{{{d}_{i}}+{{\sigma}_{i}}}.
\end{equation}

\noindent Here, $\sigma_i$ is the pseudo-gyroscale, defined as $v_{th,\parallel}/\Omega_i$ using the parallel proton temperature, $T_{i,\parallel}$, which we distinguish from the typical definition of the ion gyroscale, $\rho_i$. The waves do not necessarily need to be parallel-propagating for resonance to occur; as long as there is a large enough $k_\parallel$ component, a wave can resonate with the proton population, even if it also has a significant $k_\perp$ component. If there is a substantial population of AICs in the solar wind, then we may expect the spectral break to occur at the scale $1/k_c$.

Several past studies have explored the physical processes behind the observed spectral steepening by comparing characteristic ion scales with the measured spectral break from \textit{in-situ} data \citep{Leamon1998a,Leamon2000,Smith2001,Markovskii2008,Perri2010,Bourouaine2012,Bruno2014,Chen2014,Roberts2017DirectTurbulence,Wang2018Ion-scaleTurbulence}, or through simulations \citep[e.g.,][]{Ghosh1996SimulationMagnetohydrodynamics,Howes2008KineticPlasmas,Cerri2016Subproton-ScaleSimulations,Franci2016PlasmaSimulations,Franci2017MagneticTurbulence}. However, these studies have produced various different conclusions, and there is currently no consensus on the dominant dissipation mechanism. The difficulty in determining the break scale arises from the fact that the measured scales $d_i$ and $\rho_i$ at 1 AU are linked by the proton perpendicular plasma beta, $\beta_{i,\perp}=n_ik_BT_{i,\perp}/\left(B_0^2/2\mu_0\right)$,

\begin{equation} \label{equ:beta}
\frac{{{\rho }_{i}}}{{{d}_{i}}}=\sqrt{{{\beta}_{i,\perp}}},
\end{equation}

\noindent and typically $\beta_{i,\perp}\sim$1, so that these scales are inseparable, except in cases where $\beta_{i,\perp}\ll$1 or $\beta_{i,\perp}\gg$1 \citep[for example, see][]{Chen2014}. Therefore, the spectral break may be associated with different scales, depending on changing solar wind conditions.

\subsection{Coherent Helicity Signature at High Frequencies}

We can gain a better understanding of the possible dissipation mechanisms by looking at the nature of the fluctuations at these frequencies. The presence of fluctuations with different properties such as polarization will limit the role of certain mechanisms under different conditions. A useful quantity that can be used to diagnose certain types of fluctuations is the magnetic helicity, which characterizes the solenoidal structure of the magnetic field and twistedness of field lines \citep[][see also \citealt{Smith2003MagneticWind,Telloni2013}]{Moffat1978,Woltjer1958a}. For solar wind turbulence, the quantity of interest is the fluctuating magnetic helicity density \citep{Matthaeus1982}. A reduced form, $H_m(k)$, can be computed from single-spacecraft measurements, which based on several assumptions \citep{Batchelor1970,Matthaeus1982a,Montgomery1981}, is:

\begin{equation}
H_{m}(k)=\frac{{{2}_{{}}}\text{Im}\{\mathbf{P}_{yz}(k)\}}{k},
\end{equation}

\noindent where $\mathbf{P}_{yz}$ is the $y-z$ component of the reduced power spectral tensor of the magnetic field fluctuations in Geocentric Solar Ecliptic (GSE) coordinates \citep[for details on reduced spectra, see][]{Wicks2012}. We define the reduced normalized magnetic helicity, $\sigma_m(k)$, as:

\begin{equation} \label{equ:hel}
{{\sigma }_{m}}(k)=\frac{{{k}_{{}}}{{H}_{m}}(k)}{{{E}_{b}}(k)}\equiv \frac{{{2}_{{}}}\text{Im}\left\{ {{\mathbf{P}}_{yz}}(k) \right\}}{\text{Tr}\left\{ {{\mathbf{P}}_{ij}}(k) \right\}}.
\end{equation}

\noindent Here, $E_b(k)$ is the reduced magnetic spectral energy, which is given by the trace of the reduced power spectral tensor: $\text{Tr}\{\mathbf{P}_{ij}\}=\mathbf{P}_{xx}+\mathbf{P}_{yy}+\mathbf{P}_{zz}$. The normalized magnetic helicity gives a dimensionless measure of the polarization of magnetic fluctuations to identify wave modes at a particular frequency in the turbulent spectrum; $\sigma_m$ is zero for linearly polarized waves and $\pm$1 for right- or left-hand circularly polarized fluctuations, respectively.

Past studies using a global mean magnetic field have found a lack of coherent helicity at low frequencies in the inertial range, i.e., fluctuating almost randomly between negative and positive values \citep{Matthaeus1982}. However, at ion-kinetic frequencies there is a dominant coherent signature that suggests right-hand polarization for outward propagating fluctuations \citep{Goldstein1994,Leamon1998a,Hamilton2008,Brandenburg2011ScaleWind,Markovskii2015}. More recently, wavelet-based studies \citep[][and references therein]{Telloni2012} using the technique first developed by \citet{Horbury2008} for local mean field analysis have been employed \citep[see also,][for more details]{Podesta2009,Forman2011,Podesta2011,Wicks2010}. These studies attribute the right-handed signature to the presence of KAWs propagating at large angles to the local mean field and have also revealed the presence of a weaker left-hand polarized component due to quasi-(anti)parallel propagating AICs \citep{He2011,He2012a,He2012,Podesta2011,Klein2014,Bruno2015,Telloni2015}.

From these results, we may interpret the coherent helicity signature first observed by \citet{Goldstein1994} as arising from the dominance of the right-handed component over the left-handed component, implying dissipation of AICs at these frequencies that may be due to ion-cyclotron resonance. In fact, \citet{Bruno2015} showed that transitioning from fast to slow wind in the trailing edge of a fast wind stream (i.e., for decreasing Alfv\'enicity), both signatures weaken and eventually disappear, although, the left-handed component is first to fade completely. However, \citet{Howes2010} showed KAWs alone can also reproduce the observed helicity signature without the need for cyclotron resonance.

In this paper, we present a rigorous analysis of solar wind turbulence at ion-kinetic frequencies using a combined identification of the frequency of the spectral break and onset of the magnetic helicity signature. We compare these spectral properties of the fluctuating magnetic field with the characteristic plasma scales, $d_i$, $\rho_i$, and $1/k_c$, and attempt to link the coherent helicity signature with the spectral steepening to help identify possible dissipation mechanisms at these frequencies. We use magnetic field spectra at a much higher resolution than undertaken previously so that plasma scales do not vary considerably over the time-series of data used to compute the spectra. Our use of a large dataset over the course of a year also enables us to identify how changing solar wind conditions affect possible dissipation mechanisms. We find evidence of proton cyclotron resonance that occurs at least half the time in our studied interval, particularly in the more Alfv\'enic fast wind, and discuss the possible implications for plasma heating at ion-kinetic scales.

\section{Data Analysis and Results}

For this study, we use data from the \textit{Wind} spacecraft \citep{Acuna1995}, which launched in 1994. It moved permanently to the L1 point in 2004, providing almost 14 years of continuous \textit{in-situ} solar wind measurements. We obtain high-resolution 11 Hz (every 0.092 ms) magnetic field measurements in GSE coordinates from the MFI instrument \citep{Lepping1995}, using the calibration of \citet{Koval2013}, and ion moments at a resolution of 92 seconds, including solar wind speed, $v_{sw}$, proton density, $n_i$, and proton temperatures, $T_{i,\parallel}$ and $T_{i,\perp}$, from the SWE instrument \citep{Ogilvie1995}, using the fitting technique described by \citet{Maruca2013}. We pre-process the magnetic field data by removing small data gaps (<10 measurements, about 1 second of data) with linear interpolation, but leave larger gaps present. Similarly, we interpolate over small data gaps (<3 measurements, about 5 minutes) for the plasma moments. We also remove any plasma data from our analysis flagged as having unreliable fitting and remove manually any unphysical and anomalous measurements not identified by flagging. We use an entire year of data from 2012 in our analysis; this large dataset outweighs the presence of a small number of large data gaps, while any smaller gaps that are interpolated should have a minimal impact on our overall results.

Due to the small amplitude of the turbulent fluctuations at ion-kinetic frequencies, instrumental and spacecraft-induced noise can lead to an artificial flattening of the power spectrum at the highest frequencies. For the MFI instrument, this `noise-floor' is thought to arise from the analog-to-digital conversion of the signal, the spacecraft spin, and spin-tone harmonics. The only past measurement for the noise level of the MFI instrument was by \citet{Lepping1995} (see Figure 3(b) therein), which was conducted on a prototype sensor before launch. To ensure that the amplitudes of power spectra at high frequencies are physical, we first determine the amplitude and frequency-dependence of the MFI noise-floor from in-flight measurements before analyzing solar wind data. We provide details of this `noise-floor' determination in Appendix \ref{sec:appB} and provide this dataset for use in future studies.

\subsection{Analysis of Solar Wind Spectra} \label{sec:spectra}

To compute solar wind spectra, we employ a continuous wavelet transform (CWT) with a Morlet wavelet of frequency-width, $\omega_0=6$, using the method described by \citet{Torrence1998}. We obtain wavelet coefficients, $W(s,t)$, as functions of the scale, $s$, at which the wavelets are evaluated, and time. We then convert these scales into equivalent Fourier frequencies using $f\approx\omega_0/2\pi s$ and calculate components of the reduced power spectral tensor,

\begin{equation}
{{\mathbf{P}}_{ij}}(f,t)={{W}_{i}}(f,t)W_{j}^{*}(f,t),
\end{equation}

\noindent where the asterisk indicates complex conjugate and the indexes describe the three GSE coordinates, $i,j=x,y,z$. The power spectral density (PSD) is then:

\begin{equation} \label{equ:psd}
\text{PSD}(f,t)=\frac{2}{{{f}_{s}}}\text{Tr}\left\{ {{\mathbf{P}}_{ij}}(f,t) \right\},
\end{equation}

\noindent where $f_s=10.87$ Hz is the sampling frequency of the MFI instrument. We first pad the signal to account for any border effects arising from the finite width of the Morlet wavelet, and then calculate an estimate of the PSD at each frequency for every measurement of the original time-series. After removal of padding, we average the PSD over every 1000 MFI measurements. This averaging improves the accuracy of the amplitude of the power spectrum at each scale and results in one spectrum for every 92 s of data, which is the cadence of the SWE instrument. We take the time-stamp of each 92 s spectrum as the middle of the time-series used to produce that spectrum. Finally, we interpolate the time-series of plasma measurements onto the time-series of averaged PSD estimates to associate one measurement of the ion moments with each 92 s power spectrum.

We note that the length of time over which we average the spectra is shorter than the correlation time of solar wind turbulence, and so the assumptions of stationarity and ergodicity do not hold at low frequencies \citep{Matthaeus1982b,Perri2010a}. As such, many of the usual results of turbulence are not recovered, for example, the spectra do not converge to the typical $f^{-5/3}$ power law expected in the inertial range. However, we are attempting to measure turbulent behavior at ion-kinetic frequencies (0.1-5.5 Hz) and not in the inertial range. At these high frequencies there are a larger number of wavelengths sampled at these smaller scales during the advection of the turbulence past the spacecraft, and therefore, the stationarity and ergodicity conditions are satisfied in our dataset for our frequency range of interest.

\subsection{Estimation of the Break Frequency} \label{sec:break}

To estimate the break frequency of each spectrum, $f_b$,  we fit the PSD to the following linear function:

\begin{equation}
{{\log }_{10}}(\text{PSD})=m\,{{\log }_{10}}(f)+c,
\label{equ:fit}
\end{equation}

\noindent where $m$ is the gradient of the line or spectral exponent. To accommodate for greater uncertainty in the spectra at low frequencies, we fit this function to the power spectra using windows in frequency that increase in width in logarithmic space towards lower frequencies, giving us a value for $m$ for each window. The frequencies, $f$, for fitting Equation (\ref{equ:fit}) to the spectrum included in each window are given by:

\begin{equation}
\log_{10}{(f_m)}-0.1j\le\log_{10}{(f)}\le\log_{10}{(f_m)},
\end{equation}

\noindent where $f_m$ is the maximum frequency within the window and the index, $j=1,2,3,...$, increases by an integer factor for each successive window so that the term $0.1j$ widens the window as $j$ increases. For each successive window, we set:

\begin{equation}
\log_{10}{(f_{m,j+1})}=\log_{10}{(f_{m,j})}-0.1j/20,
\end{equation}

\noindent shifting the windows to lower frequencies as $j$ increases. The division by 20 in the last term allows us to overlap the windows and provide a sufficient number of fits for $m$ over the frequencies at which we evaluate the power spectra. We continue our windowing process along the spectrum as long as $\log_{10}{(f_m)}>-1$, giving us a total of 26 windows. The center frequency of each window, which we associate with a value of $m$, is taken as the median of the frequencies in that window.

\begin{figure}
\centering
\includegraphics[width=0.425\textwidth]{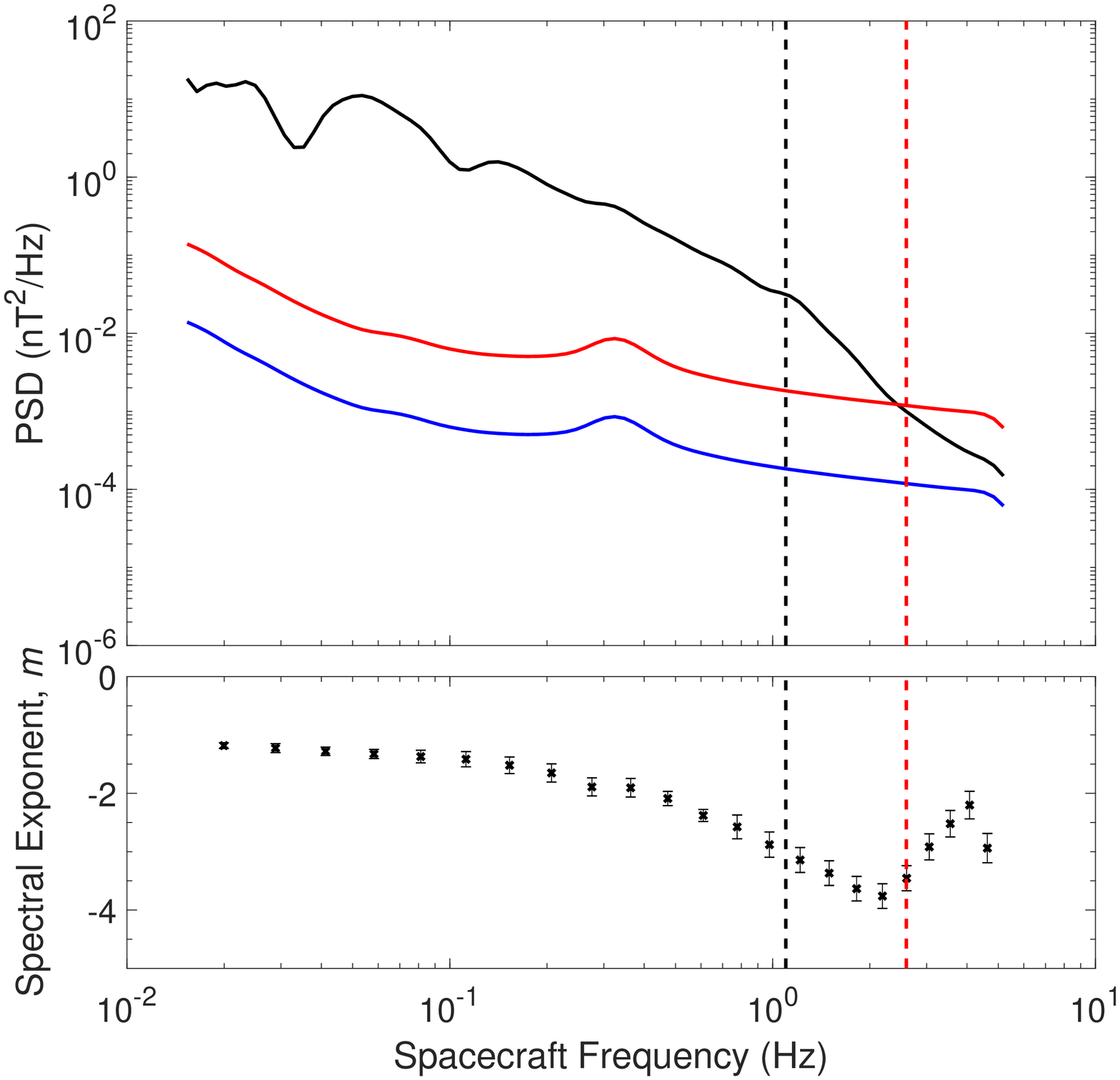} 
\caption{$Top$: An example 92 s solar wind magnetic field power spectrum, in black. The blue line is the MFI noise-floor from Appendix \ref{sec:appB}, the red line is the noise-floor multiplied by a signal-to-noise ratio of 10, and the red-dashed line is the noise cut-off frequency, $f_{noise}$ (see main text). $Bottom$: Results from the fitting of the function (Equation \ref{equ:fit}) to the spectrum, showing the spectral exponent, $m$, for each window in our fitting process. Error bars show the root-mean-square error of the fitting. The black dashed-line is our estimated break frequency, $f_b$.}
\label{fig:1}
\end{figure}

\begin{figure*}
\begin{center}
\includegraphics[scale=0.35]{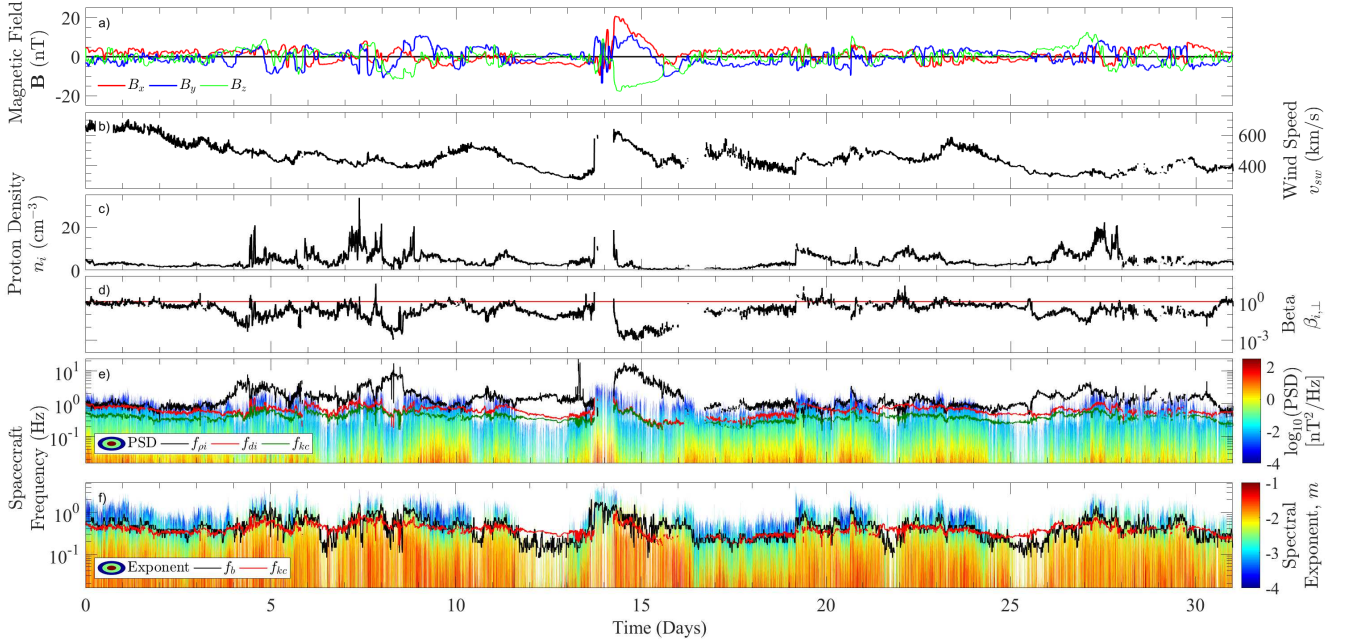}
\caption{July 2012 time series of (a) the components of the magnetic field, $\bf{B}$, smoothed using a 51-point median filter, (b) the solar wind speed, $v_{sw}$, (c) the solar wind proton density, $n_i$, and (d) the proton perpendicular plasma beta, $\beta_{i,\perp}$. In panel (d), the red line indicates $\beta_{i,\perp}=1$, where $\rho_i=d_i$ and therefore $1/k_c\simeq\rho_i+d_i=2\rho_i=2d_i$, from Equations (\ref{equ:kc}) and (\ref{equ:beta}), assuming $\sigma_i\simeq\rho_i$. (e) Contour plot of consecutive 92 s solar wind magnetic field spectra. The white areas indicate large data gaps or data with frequencies $f\geq f_{noise}$. We show the characteristic plasma scales, $1/k_c$, $d_i$, and $\rho_i$, converted to frequencies using Taylor's hypothesis (see main text) as the solid green, red, and black lines, respectively. (f) Contour plot of the spectral exponent, $m$, to the corresponding power spectra in panel (e). We also plot $f_{kc}$ in red and the estimated $f_b$ in black for comparison, which we smooth here by a 21-point median filter to improve visualization of the plot.}
\label{fig:2}
\end{center}
\end{figure*}

We show in the top panel of Figure 1 an example 92 s spectrum from July 2012 and in the bottom panel, our results for $m$ from our fitting process. We see a change in $m$ from about -1.2 to -3.8 from low to high frequencies, resulting from the transition between the power laws for the inertial range and the ion-kinetic range. This transition is not a simple step-function because of the finite width of the fitting window at the frequency of the spectral break. To determine the width of this transition, we calculate two frequencies that bound either side of it. We identify the first, $f_1$, when the difference between two successive values for $m$ exceeds the threshold $\left|m_{j+1}-m_j\right|\ge0.05$, and the second, $f_2$, as the frequency with the minimum value for $m$. We then estimate the break frequency $f_b$ for each spectrum as $f_b=(f_1+f_2)/2$, in a similar fashion to \citet{Chen2014}. The black dashed-line in Figure 1 is our estimate of $f_b$ for the example spectrum using this method, which we see agrees well with the break in the spectrum.

Towards higher frequencies in Figure 1, the spectrum flattens and $m$ increases. This flattening is most likely due to the increasing contribution of instrumental noise to the signal at these frequencies. To ensure that our estimated $f_b$ is physical, we determine a cut-off frequency, $f_{noise}$, where the spectrum is equal to a signal-to-noise ratio (SNR) of 10 times our noise-floor estimate (see Appendix \ref{sec:appB}), indicated by the vertical red dashed-line in Figure 1. We neglect an estimate of the break frequency if $f_b\geq f_{noise}$. Close to the Nyquist frequency, there is a second decrease in the spectral exponent, which we attribute to artifacts of the CWT.

To test the robustness of our automated fitting procedure and method to calculate $f_b$, we first apply it to consecutive 92 s power spectra over the course of one month, using data from July 2012. Panels (a-d) in Figure 2 show time-series of the components of the magnetic field, $\bf{B}$, the solar wind speed, $v_{sw}$, proton density, $n_i$, and proton perpendicular beta, $\beta_{i,\perp}$, respectively. We smooth $\bf{B}$ using a 51-point median filter here for visual purposes to emphasize the sectoral structure of the interplanetary magnetic field from the numerous crossings of the heliospheric current sheet, highlighted by the changing sign of the $B_x$ and $B_y$ components. We see that $v_{sw}$ varies between 300 and 700 km/s and $n_i$ from less than 1 cm$^{-3}$ to almost 35 cm$^{-3}$. There are several periods, often during fast wind intervals, where $\beta_{i,\perp}\sim1$. At other times $\beta_{i,\perp}$ typically does not exceed unity and reaches a minimum value of almost $1\times10^{-3}$. The spacecraft sampled periods of both slow and fast wind, as well as shocks, density enhancements, and transient ejecta, illustrating the variability of the solar wind during this interval.

Panel (e) in Figure 2 shows a contour plot of consecutive 92 s power spectra over July 2012, i.e., a time series of spectra over the course of a month. In comparison with panels (a-d), we see that the spectra and therefore, the turbulent processes in the solar wind, depend on the overall plasma conditions, particularly at high frequencies. Here, white areas indicate data that we have removed, either due to the presence of a large data gap or because the frequencies exceed the defined noise-floor cut-off, $f_{noise}$. We also show as solid lines the three characteristic plasma scales, $1/k_c$, $d_i$, and $\rho_i$, in green, red, and black, respectively. We plot these three scales as frequencies assuming Taylor's hypothesis: $f_L=v_{sw}/2\pi l$, where $l$ is the appropriate length scale. According to panel (e), there are several periods during which $f_{noise}<f_L$, emphasizing the importance of our noise-floor treatment.

In Figure 2(f) we show a contour plot of the spectral exponent, $m$, versus frequency for each corresponding spectrum in panel (e), along with our estimated $f_b$ and $f_{kc}$ in black and red, respectively, for comparison. We note that the break frequencies are discretized by the scales of the wavelets and hence, the windowing process in our fitting procedure. We discard values where $f_b\geq f_{noise}$, and also $f_b\leq0.1$ Hz. This second condition allows us to avoid times when the amplitude of fluctuations is so low that a physical break between two power laws is obscured by noise, and therefore, an estimate for $f_b$ by our automated method is unreliable. We smooth $f_b$ here only for this Figure using a 21-point median filter. We find that our fitting procedure performs an accurate estimate of $f_b$ for the $\sim$29,000 spectra from July 2012, since $f_b$ agrees well with the break in the spectrum from visual inspection of panel (f).

\begin{figure*}
\begin{center}
\includegraphics[scale=0.8]{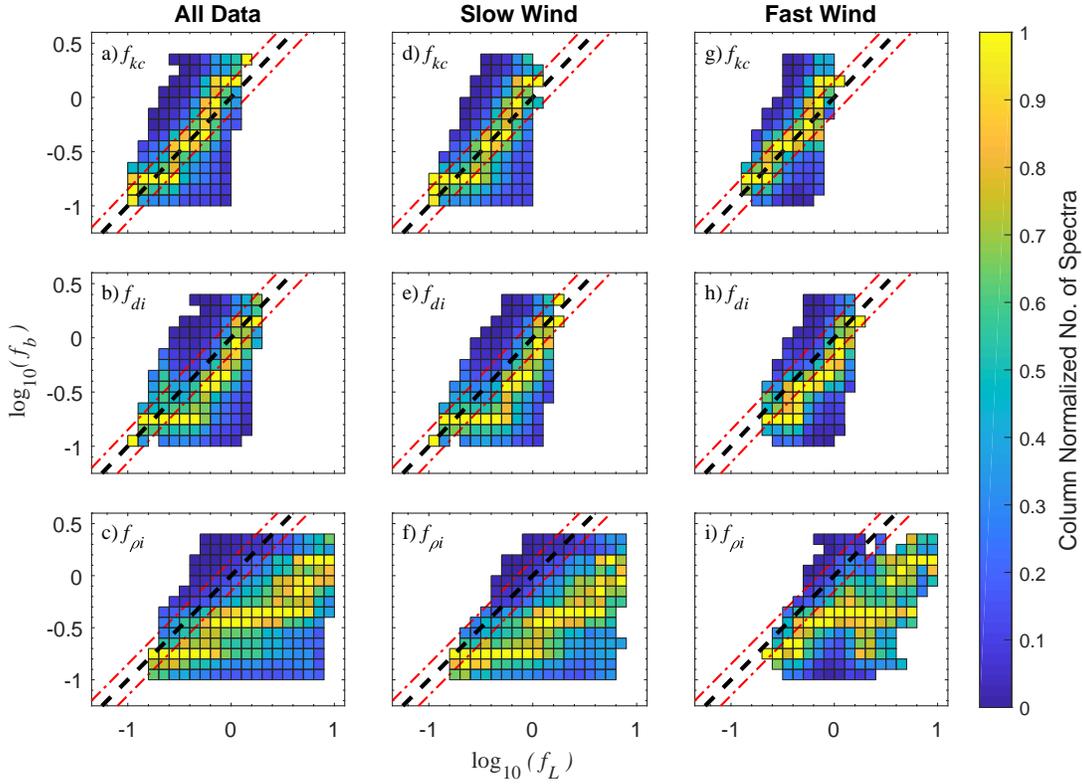}
\caption{(a-c) Histograms for 2012 of the estimated break frequency, $f_b$, versus the three characteristic plasma scales, converted into frequencies using Taylor's hypothesis - $f_L$ represents $f_{kc}$, $f_{di}$ and $f_{\rho i}$, for each row respectively. (d-f) The corresponding results for only slow wind (<400 km/s) intervals. (g-i) The corresponding results for only fast wind (>500 km/s) streams. The color-bar represents the column-normalized number of spectra. The black dashed lines represent $f_b=f_L$ and similarly, the red dashed lines are $f_b=f_L\;\sqrt[]{2}$ and $f_b=f_L/\,\sqrt[]{2}$, which give the resolution of the wavelet transform about the line $f_b=f_L$.}
\label{fig:3}
\end{center}
\end{figure*}

We now compare the three plasma scales as frequencies, $f_L$, with $f_b$, where $L=1/k_c$, $d_i$, and $\rho_i$, and extend our analysis to a year of data from 2012. We calculate $\sim$ 344,000 spectra, estimating $f_b$ for each spectrum and compare it to the corresponding values for the characteristic plasma scales, $f_L$. Figure 3 shows two-dimensional histograms for $f_b$ with $f_{kc}$, $f_{di}$, and $f_{\rho i}$ in the top, middle, and bottom rows, respectively. We show the results for all data and then separate according to slow ($v_{sw}$<400 $\text{km s}^{-1}$) and fast wind ($v_{sw}$>500 $\text{km s}^{-1}$) in the left, middle, and right columns, respectively. Separating by wind speed allows us to test for systematic effects due to large-scale solar wind stream structure. We normalize each column of the binned data in each plot by the maximum number of spectra in a bin for that column, highlighting the most probable $f_b$ measured as a function of $f_L$. We neglect values for $f_b$ when $f_b\geq f_{noise}$ and $f_b\leq0.1$ Hz, and to avoid under-sampling. We also omit bins with $\leq$10 spectra to avoid under-sampling. In each panel, the black dashed lines give the line $f_b=f_L$ and similarly, the red dashed lines are $f_b=f_L\;\sqrt[]{2}$ and $f_b=f_L/\,\sqrt[]{2}$, which indicate the resolution of the wavelet transform about the line $f_b=f_L$ due to the finite width of the Morlet wavelet in frequency space \citep[i.e., the \textit{e}-folding frequency, see][]{Torrence1998}.

To quantify any relationship between $f_b$ and $f_L$, we conduct a statistical analysis using this year of data. We first calculate the Pearson correlation coefficient,

\begin{equation} \label{equ:coeff}
R(f_b,f_L)=\frac{1}{N-1}\sum\limits_{{i=1}}^{{N}}{{\left(\frac{f_{b,i}-\mu_b}{\sigma_b}\right)\left(\frac{f_{L,i}-\mu_L}{\sigma_L}\right)}},
\end{equation}

\noindent where $\mu$ is the mean and $\sigma$ is the standard deviation. The coefficient $R\in\left[-1,+1\right]$ measures the linear correlation between $f_b$ and $f_L$. A value of $R=\pm1$ indicates a positive or negative linear correlation, respectively, whereas zero indicates no linear correlation. If $\left|R\right|=1$, a linear equation describes the relationship between the variables $f_b$ and $f_L$. We also define a residual, $\rho$, which in analogy to the standard deviation is:

\begin{equation} \label{equ:res}
\rho(f_b,f_L)=\sqrt[]{\frac{1}{N-1}\sum\limits_{{i=1}}^{{N}}{{\left|f_{b,i}-f_{L,i}\right|}^2}},
\end{equation}

\begin{figure*}
\centering
\includegraphics[scale=0.8]{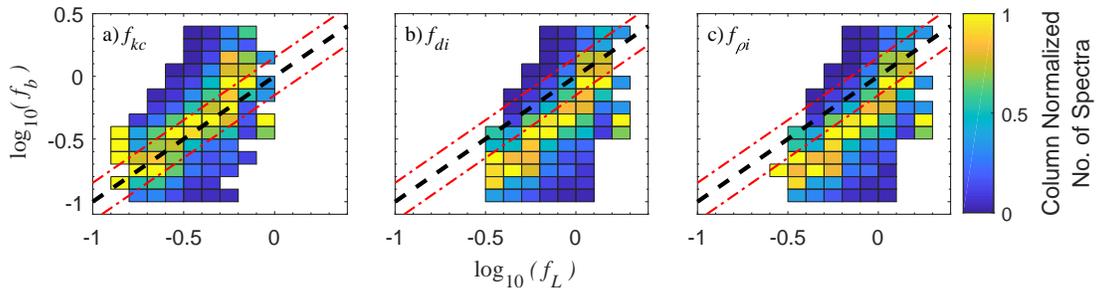}
\caption{Histograms for 2012 of the estimated break frequency, $f_b$, versus the three characteristic plasma scales, converted into frequencies using Taylor's hypothesis - $f_L$ represents $f_{kc}$, $f_{di}$ and $f_{\rho i}$, for each column respectively. The data used are for periods where $0.95\geq\beta_{i,\perp}\leq1.05$. The color-bar represents the column-normalized number of spectra. The black dashed lines represent $f_b=f_L$ and similarly, the red dashed lines are $f_b=f_L\;\sqrt[]{2}$ and $f_b=f_L/\,\sqrt[]{2}$, which give the resolution of the wavelet transform about the line $f_b=f_L$.}
\label{fig:4}
\end{figure*}

\noindent where $\rho\geq0$. The residual gives the difference between our measured $f_L$ and estimated $f_b$, in other words, a value of $\rho$ closer to zero indicates there is less spread of data about the line $f_b=f_L$. In our calculations of $R$ and $\rho$, we take the logarithms of both $f_b$ and $f_L$. We place more weight here on the statistical significance of $\rho$ over $R$ since a linear relationship does not necessarily imply that $f_b=f_L$. The correlation coefficients are also intrinsically linked because of the similar definitions of the three plasma scales in Equations (\ref{equ:kc}) and (\ref{equ:beta}), especially when $\beta_{i,\perp}\sim1$. As a final test, we count the number of spectra that lie between the two red dashed lines in each panel of Figure 3 to determine a percentage of the total number of spectra that satisfy $f_b\simeq f_L$, within the \textit{e}-folding frequency. For the total number of spectra, we do not include instances where there are large data gaps, and where we have discarded values for $f_b$. For instances where we filter data according to slow or fast wind, the total number of spectra we use is only for that filtered dataset. We give the results of our statistical analysis in Table \ref{tab:1}.

\begin{deluxetable}{lcccc}[b!]
\tablecaption{Correlation coefficients, residuals, and percentages for $f_L$ and $f_b$ from the data shown in Figures 3 and 4.\label{tab:1}}
\tablehead{
\colhead{} &
\colhead{Plasma Scale} &
\colhead{Correlation} &
\colhead{Residual} &
\colhead{Percent} \\
\colhead{} &
\colhead{$L$} &
\colhead{$R$} &
\colhead{$\rho$} &
\colhead{\%}
}
\startdata
& $k_c$ & 0.56 & 0.17 & 52.08 \\
All Data & $d_i$ & 0.52 & 0.25 & 30.19 \\
& $\rho_i$ & 0.34 & 0.43 & 9.14 \\
\hline
& $k_c$ & 0.54 & 0.12 & 49.42 \\
Slow & $d_i$ & 0.48 & 0.19 & 27.74 \\
& $\rho_i$ & 0.38 & 0.33 & 6.41 \\
\hline
& $k_c$ & 0.58 & 0.06 & 57.01 \\
Fast & $d_i$ & 0.58 & 0.08 & 36.08 \\
& $\rho i$ & 0.32 & 0.14 & 15.66 \\
\hline
& $k_c$ & 0.60 & 0.02 & 51.81 \\
$\beta_{i,\perp}\sim1$ & $d_i$ & 0.61 & 0.04 & 26.44 \\
& $\rho_i$ & 0.61 & 0.04 & 26.42 \\
\vspace{-0.38cm}
\enddata
\end{deluxetable}

For all 2012 data, regardless of wind speed, both $f_{kc}$ and $f_{di}$ have moderate correlations with $f_b$, with values of $R=0.56$ and $R=0.52$, respectively, whereas the correlation for $f_{\rho i}$ is weaker at $R=0.34$. The lowest residual is $\rho=0.17$ for $f_{kc}$, while for $f_{di}$ it is $\rho=0.25$ and for $f_{\rho i}$ it is even higher at $\rho=0.43$. These values show that the cyclotron resonance scale, $1/k_c$, is most closely associated with the spectral break (i.e., closest to $f_b\simeq f_L$) during the interval we study. This finding is supported by 52.08\% of the total number of spectra in our dataset falling within the two red dashed-lines for $f_{kc}$ in Figure 3(a). In contrast, we find only 30.19\% for $f_{di}$ in panel (b) and 9.14\% for $f_{\rho i}$ in panel (c). The gyroscale, $\rho_i$, therefore has a poor relationship with the break frequency, suggesting that it is least likely to be associated with the spectral steepening during the interval we study.

These findings hold when we separate the data according to wind speed. For slow wind, $f_{kc}$ has the highest correlation coefficient at $R=0.54$ and lowest residual at $\rho=0.12$. During periods of fast wind streams, the residual for $f_{kc}$ is about half that of slow wind at $\rho=0.06$, the smallest for all three scales. From panels (d) and (g) in Figure 3, we are unable to visually differentiate between the fast and slow wind cases without statistical analysis. Also, comparing panels (f) and (g), we find that 57.01\% of spectra fall within the resolution limit of our wavelet transform for $f_{kc}$ in fast wind, which is higher than for slow wind at 49.42\%. The correlation coefficients for both $f_{kc}$ and $f_{di}$ in fast wind are equal, at $R=0.58$, which are only slightly larger than their slow wind values. Transitioning from slow to fast wind in panels (e) and (h), the percentages of spectra where $f_b\simeq f_{di}$ increase from 27.74\% to 36.08\%. From these values, we see that the relationship between $f_{kc}$ and $f_b$ is maintained even when the large-scale stream structure of the wind varies but is strongest in fast wind streams.

According to Equations (\ref{equ:kc}) and (\ref{equ:beta}), $1/k_c$ will coincide with the larger of the two scales, $d_i$ or $\rho_i$, when $\beta_{i,\perp}\ll1$ or $\beta_{i,\perp}\gg1$, respectively, assuming an isotropic temperature (i.e., $\rho_i\simeq\sigma_i$). This expectation is consistent with observations by \citet{Chen2014} showing that the spectral break occurs at $d_i$ for $\beta_{i,\perp}\ll1$ and at $\rho_i$ for $\beta_{i,\perp}\gg1$, which they note is consistent with a break at $1/k_c$ in both cases. However, by definition, when $\beta_{i,\perp}\sim$1, $\rho_i\simeq d_i$ and therefore, $1/k_c\simeq\rho_i+d_i\simeq 2\rho_i\simeq 2d_i$. For periods with $\beta_{i,\perp}\sim1$ as seen in Figure 2(d-f), both $f_{di}$ and $f_{\rho i}$ coincide and $f_{kc}$ is shifted to lower frequencies by about a factor of 2. During these periods, there is a good agreement between $f_{kc}$ and $f_b$. To address what happens when $\beta_{i,\perp}\sim1$ quantitatively and clearly show the difference between $1/k_c$ and $d_i$ or $\rho_i$, we filter our year of data to include only periods where $0.95\geq\beta_{i,\perp}\leq1.05$ and show the corresponding 2D histograms in Figure 4. In addition, the results from our statistical analysis are shown in the bottom panel of Table \ref{tab:1}. We note that we do not remove bins with $\leq$10 spectra here due to the smaller amount of data available for these periods, but this only affects bins furthest from the black dashed-line.

Comparing panels (b) and (c) to (a) in Figure 4 we see that our measured $f_b$ is consistently shifted to frequencies lower than $f_{di}$ and $f_{\rho i}$, i.e., the yellow enhancement in panels (b) and (c) is below the black dashed-line, but in panel (a) we see that it is closer to the dashed-line. These plots show that $1/k_c$ is a more likely candidate for the break scale than $d_i$ or $\rho_i$, and we quantify this result by calculating $R$ and $\rho$ for this dataset. The correlation coefficients are the same for $f_{di}$ and $f_{\rho i}$ at $R=0.61$, and almost the same at $R=0.60$ for $f_{kc}$, however, the latter has the lowest residual at $\rho=0.02$, compared to $\rho=0.04$ for $f_{di}$ and $f_{\rho i}$. We note that the statistics for $f_{\rho i}$ improve considerably when considering only periods of $\beta_{i,\perp}\sim$1, and are the same in this case for $f_{di}$. Again, we find a high percentage of the number of spectra where $f_b\simeq f_{kc}$ at 51.81\%, almost double that for the other two scales.

When we consider all data in our interval, the results from our statistical analysis for both $f_{kc}$ and $f_{di}$ do not differ significantly, particularly in fast wind, as we see from similar values for $R$ and $\rho$ in Table \ref{tab:1}. From our analysis of periods where $\beta_{i,\perp}\sim1$, we explain this result as being due to the ratio of spectra where $\beta_{i,\perp}<$1 to $\beta_{i,\perp}>1$, which is almost 8 in our dataset. Finally, we conclude that $f_b$ is best associated with $f_{kc}$ and so the spectral break is most likely related to proton-cyclotron resonance. We then explain the correlations with $d_i$ and $\rho_i$ as due to the dependence of $1/k_c$ on both variables, and the fact that $d_i$ and $\rho_i$ are separated only by a factor of $\sqrt[]{\beta_{i,\perp}}$.

\subsection{Quantification of the Helicity Signature}

To further explore the possible role of proton-cyclotron resonance at ion-kinetic frequencies, we now investigate the nature of the fluctuations at these frequencies. We calculate helicity spectra from successive periods of 92 seconds using the normalized magnetic helicity, $\sigma_m$, from Equation (\ref{equ:hel}). Again, we use Taylor's hypothesis to obtain $\sigma_m$ as a function of frequency instead of wavenumber, giving us one helicity spectrum for each corresponding power spectrum from the previous section. To quantify the relationship between $1/k_c$ and the coherent helicity signature at high frequencies, we devise a method to calculate the helicity signature onset frequency, $f_h$, defined as the threshold frequency at which we see an enhancement in the helicity at ion-kinetic scales. We first fit a Gaussian function to the helicity spectra,

\begin{equation} \label{equ:gaussfit}
\sigma_m=\frac{1}{\sqrt[]{2\pi}\sigma_{D}}\exp{\left\{-\frac{\left(f-f_p\right)^2}{2\sigma_{D}^2}\right\}},
\end{equation}

\noindent where the fitting parameters are the standard deviation, $\sigma_{D}$, and the mean, $f_p$, which corresponds to the frequency of the peak in the helicity signature. We perform the Gaussian fitting in linear space, so that the method is biased towards the peak in helicity at the highest frequencies, i.e., the coherent helicity signature. We show in the top panel of Figure 5 the example power spectrum from Figure 1, along with its corresponding helicity spectrum in the bottom panel, both in black. In the bottom panel, we also plot in red the Gaussian fit to the helicity spectrum using Equation (\ref{equ:gaussfit}). The red dashed-line gives $f_p$ from the fitting, whereas the black and gray dashed-lines are $f_b$ and $f_{noise}$ from before, respectively. To estimate the onset frequency $f_h$, we calculate the full-width at half-maximum of the Gaussian peak using $\Delta f=\sigma_{D}\sqrt{8\ln \left( 2 \right)}$ and then,

\begin{equation} \label{equ:fh}
{{f}_{h}}={{f}_{p}}-\Delta f/2.
\end{equation}

\noindent The minus sign is used to determine the onset frequency bounded towards lower frequencies. This method is independent of whether the peak in helicity is negative or positive and allows for an automated process estimating both $f_h$ and $f_p$ for $\sim$ 344,000 helicity spectra. In Figure 5, $f_h$ is given by the blue dashed-line. We see that $f_b$ and $f_h$ are separated by only about 0.1 Hz. From the results of Section \ref{sec:break}, this result implies that $f_h$ may also be associated with $f_{kc}$ and suggests that the presence of the helicity signature is related to cyclotron resonance. We further investigate this relationship between $f_b$ and $f_h$ using the same statistical analysis from the previous section. We also follow up the work of \citet{Markovskii2015} to confirm a relationship between $f_p$ and $f_{\rho i}$ using our dataset.

\begin{figure}
\begin{center}
\includegraphics[width=0.425\textwidth]{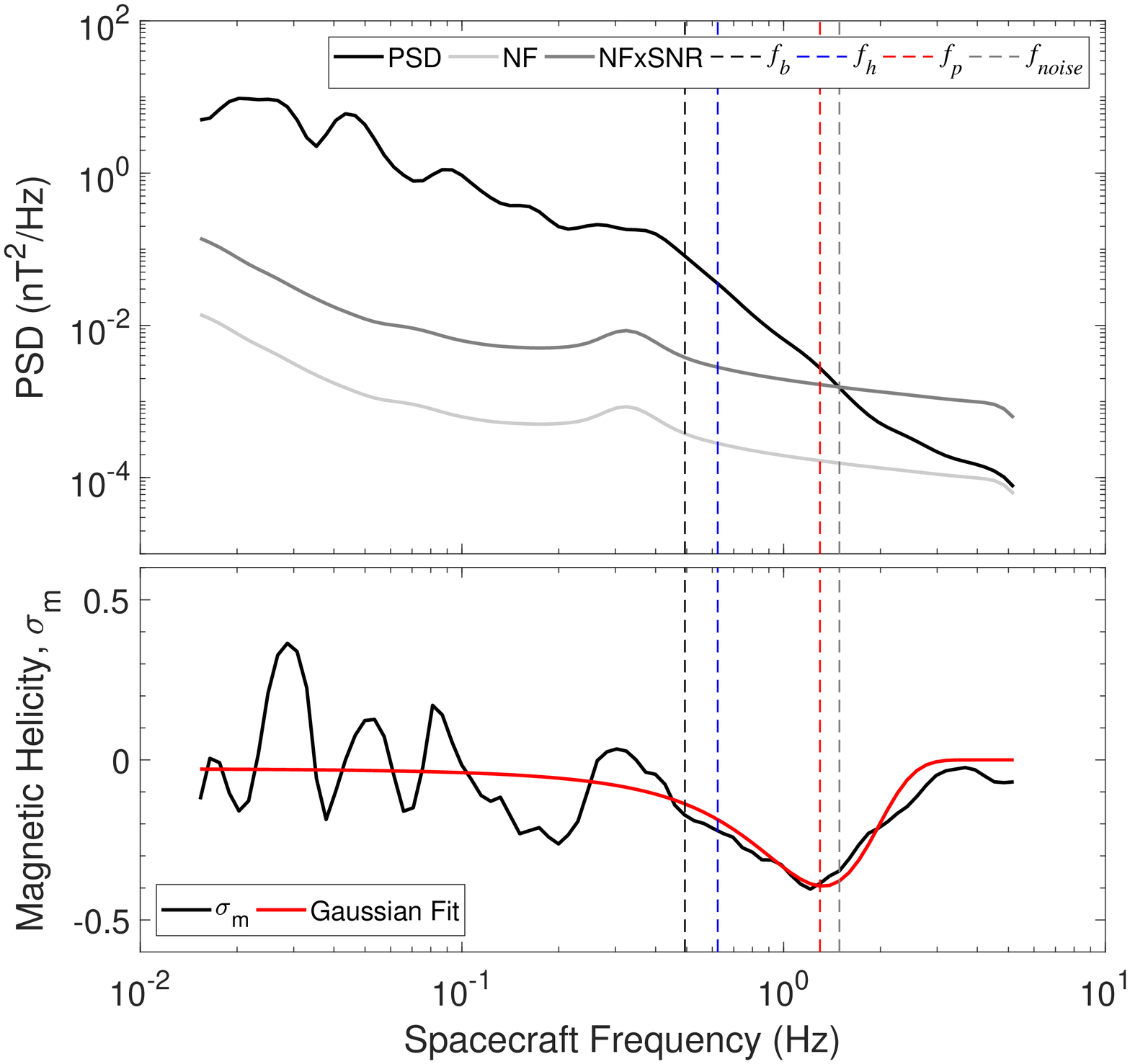}
\caption{$Top$: An example 92 s solar wind magnetic field power spectrum from Figure 1, in black. The light gray line is the MFI noise-floor from Appendix \ref{sec:appB}, the dark gray line is the noise-floor multiplied by a signal-to-noise ratio of 10, and the gray dashed-line is the noise cut-off frequency, $f_{noise}$ (see main text). The black dashed-line is our estimated break frequency, $f_b$, from before. $Bottom$: The corresponding 92 s helicity spectrum in black and the fitting of the Gaussian function (Equation \ref{equ:gaussfit}) to the spectrum in red. The coherent helicity signature at high frequencies is well-represented by the Gaussian peak. From our fitting, we obtain the helicity onset frequency, $f_h$, from Equation (\ref{equ:fh}) and the peak helicity frequency, $f_p$, from the mean of the Gaussian peak, given by the blue and red dashed-lines, respectively.}
\label{fig:5}
\end{center}
\end{figure}

\begin{figure*}
\begin{center}
\includegraphics[scale=0.35]{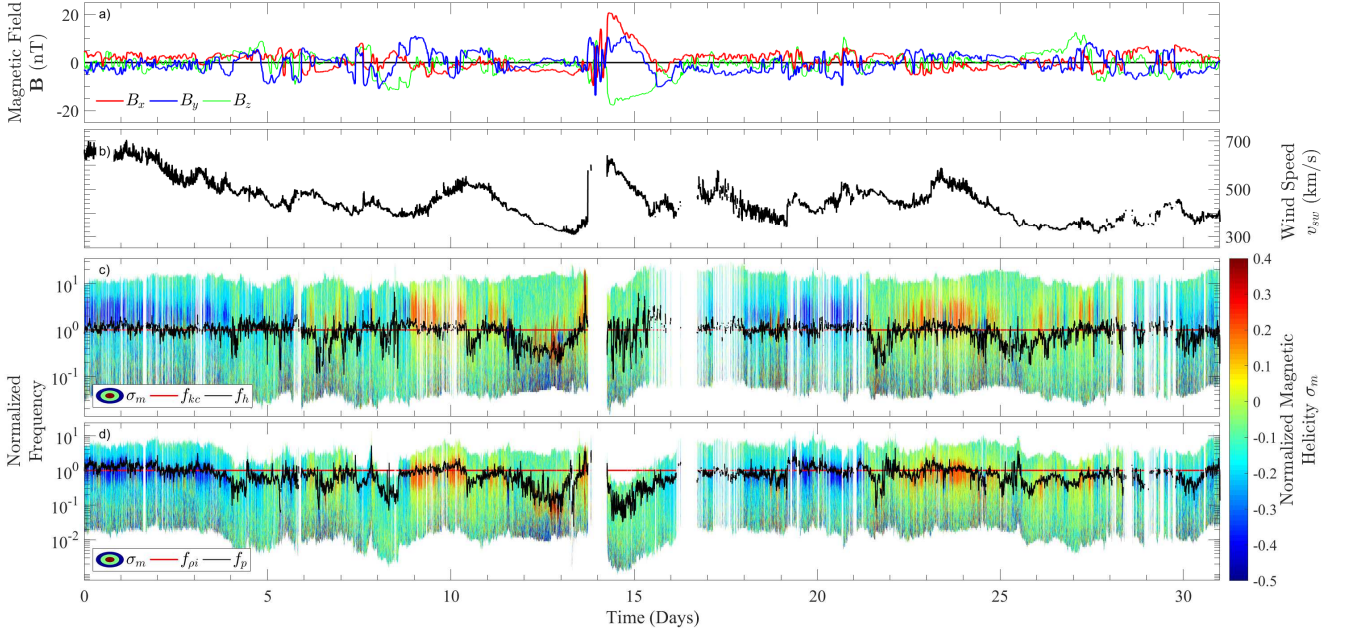}
\caption{July 2012 time series of (a) the components of the magnetic field, $\bf{B}$, smoothed using a 51-point median filter to highlight the sectoral structure of the solar wind, and (b) the solar wind speed, $v_{sw}$, repeated from Figure 2. (c) Contour plot of consecutive 92 s reduced normalized magnetic helicity, $\sigma_m$, from July 2012, corresponding to each power spectrum in Figure 2(e). The spectra have been normalized by $1/f_{kc}$, and we plot the line $f_{kc}=1$ in red for reference. We also show the estimated helicity onset frequency, $f_h$, in black. (d) Similarly to panel (c), where the spectra are normalized instead by $1/f_{\rho i}$, and we plot the line $f_{\rho i}=1$ in red for reference. We also show the estimated helicity peak frequency, $f_p$, in black. Both $f_h$ and $f_p$ are smoothed here by a 21-point median filter to improve visualization of the plot.}
\label{fig:6}
\end{center}
\end{figure*}

We first use our method described above to automatically estimate both $f_h$ and $f_p$ for July 2012 to check that it accurately reproduces the features of the helicity spectra. In Figure 6(a-b) we show again time series of $\bf{B}$ and $v_{sw}$ for July 2012. In addition, panels (c) and (d) show contour plots of $\sigma_m$ for consecutive 92 s spectra over the course of July 2012, where we normalize the frequency of each spectrum by $1/f_{kc}$ in panel (c), and similarly by $1/f_{\rho i}$ in panel (d). We also plot the lines $f_{kc}=1$ and $f_{\rho i}=1$ for reference in red, as well as our estimated $f_h$ and $f_p$ in black, in panels (c) and (d), respectively. We normalize the spectra here only for the Figure, and not in any future analysis. From panel (c), we can see that $f_h$ bounds at lower frequencies the enhanced red and blue signature as we expect. Also, from panel (d) we see that the peak of the helicity signature is located close to the middle of the helicity signature before the enhancement disappears completely (it should not be located directly in the middle due to the logarithmic scale in frequency). Therefore, we conclude that our method works as required, quantifying the helicity signature accurately. We will now discuss our findings from Figure 6 in more detail.

The persistent band of enhancement in $\sigma_m$ at higher frequencies varies between about -0.4 and +0.2. Figure 6(a) suggests that the sectoral structure of the solar wind is likely responsible for this changing sign in helicity over the course of the month, which is consistent also with the findings of \citet{He2011}. From Figure 6(c), we can see that the positive helicity signal is typically weaker in amplitude than the negative signal by a factor of 2. We currently have no explanation for this finding, but it is an interesting observation that should be explored in another study. We see that when an enhancement in helicity signature is present at high frequencies, $f_h$ is well-correlated with $f_{kc}$. By comparing panels (b) and (c), we find that the helicity enhancement weakens or almost completely disappears during periods of slow solar wind. Both these results show that the findings of \citet{Bruno2015,Telloni2015} apply to large volumes of the solar wind. Also, from Figure 6(d), the peak of this coherent helicity signature is correlated with $f_{\rho i}$, especially in fast wind where the helicity signature is strongest, which is consistent with \citet{Markovskii2015} and \citet{Telloni2015}. While we have not shown a similar plot for $f_{di}=1$, we find that it is not closely associated with either $f_h$ or $f_p$, and confirm this in our subsequent analysis.

At lower frequencies than $f_{kc}$, the helicity fluctuates about zero, as expected for the inertial range of solar wind turbulence \citep{Matthaeus1982}, showing either a lack of or no dominant coherent circular polarization of fluctuations. There is an enhanced signature in the helicity that significantly deviates from the characteristic plasma scales between the 12th and 16th July, peaking at around 0.1 Hz (from Figure 6(c), about 0.5 in normalized frequency units). We associate this signature with AICs produced by instabilities from unstable particle distributions. These waves are often Doppler-shifted towards lower frequencies than the spectral break since they typically propagate towards the Sun, in the opposite direction to the turbulent magnetic fluctuations we consider here \citep[e.g., see][]{Tsurutani1994ElectromagneticObservations,Jian2009,Jian2010ObservationsAU,Jian2014,Roberts2015,Roberts2015a,Gary2015,Wicks2016}. To exclude these events from our analysis, we discard data with $f_h\leq$0.2 Hz and $f_p\leq$0.2 Hz.

\begin{figure*}
\centering
\includegraphics[scale=0.8]{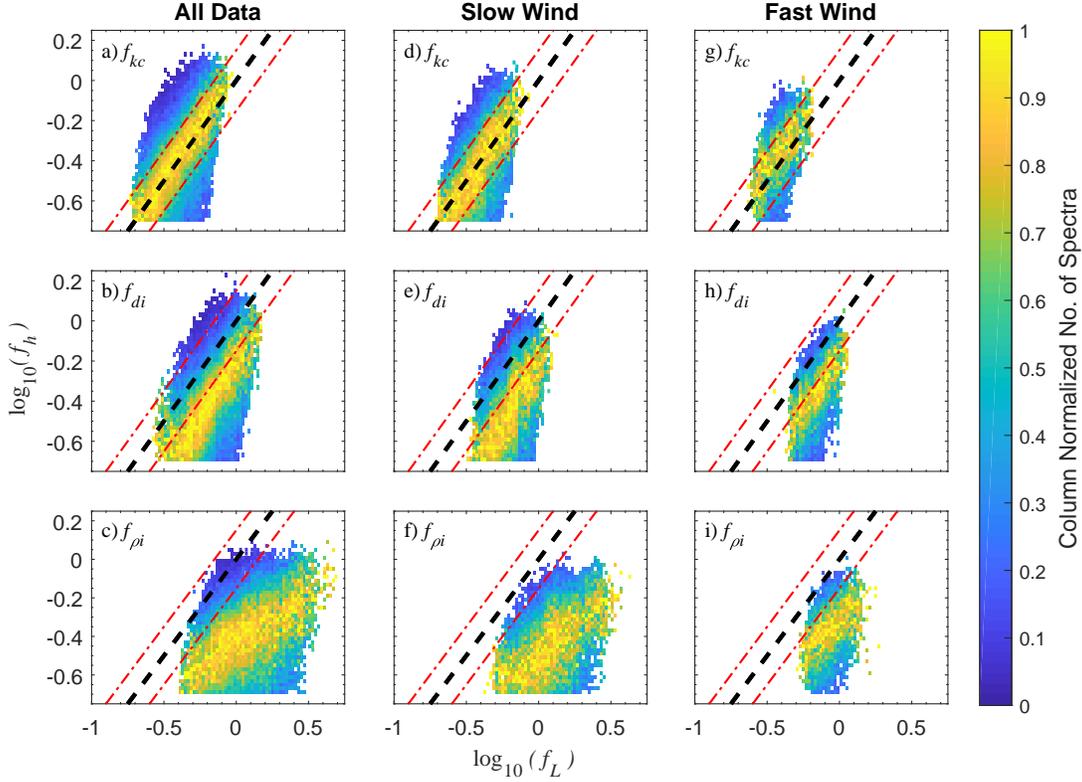}
\caption{(a-c) Histograms for 2012 of the estimated helicity onset frequency, $f_h$, versus the three characteristic plasma scales, converted into frequencies using Taylor's hypothesis - $f_L$ represents $f_{kc}$, $f_{di}$ and $f_{\rho i}$, for each row respectively. (d-f) The corresponding results for only slow wind (<400 km/s) periods. (g-i) The corresponding results for only fast wind (>500 km/s) streams. The color-bar represents the column-normalized number of spectra. The black dashed lines represent $f_h=f_L$ and similarly, the red dashed lines are $f_b=f_L\;\sqrt[]{2}$ and $f_b=f_L/\,\sqrt[]{2}$, which give the resolution of the wavelet transform about the line $f_b=f_L$.}
\label{fig:7}
\end{figure*}

At frequencies $f>f_p$, the enhancement disappears, and the helicity returns to a value close to zero. If the trace of the power spectral tensor is increased artificially by instrumental noise, then the helicity will also reduce to zero due to the influence of noise, by definition from Equation (\ref{equ:hel}). In fact, the increasing contribution from noise should not affect the phase contribution to the helicity, but rather just its amplitude. Therefore, despite seeing a return to zero, we do not see the return of the signature to an incoherent one similar to that at low frequencies seen in Figure 6 where $\sigma_m$ oscillates in color between red and blue for opposite polarizations. We do not observe the signal to fluctuate about zero, but rather remain coherent with a value close to zero for $f>f_p$. Therefore, we cannot determine whether this effect is due to the MFI noise-floor or a physical effect. An alternative explanation is aliasing (See Appendix \ref{sec:appB}). Despite this, we find that typically $f_p<f_{noise}$ and therefore take the peak in the helicity signature and hence, $f_p$, as physical.

\begin{deluxetable}{lcccc}[b!]
\tablecaption{Correlation coefficients, residuals, and percentages for $f_L$ and $f_h$ from the data shown in Figures 7 and 8.\label{tab:3}}
\tablehead{
\colhead{} &
\colhead{Plasma Scale} &
\colhead{Correlation} &
\colhead{Residual} &
\colhead{Percent} \\
\colhead{} &
\colhead{$L$} &
\colhead{$R$} &
\colhead{$\rho$} &
\colhead{\%}
}
\startdata
& $k_c$ & 0.48 & 0.09 & 64.29 \\
All Data & $d_i$ & 0.40 & 0.15 & 33.37 \\
& $\rho_i$ & 0.35 & 0.29 & 7.61 \\
\hline
& $k_c$ & 0.48 & 0.06 & 64.72 \\
Slow & $d_i$ & 0.39 & 0.10 & 32.71 \\
& $\rho_i$ & 0.38 & 0.21 & 4.97 \\
\hline
& $k_c$ & 0.45 & 0.04 & 61.38 \\
Fast & $d_i$ & 0.42 & 0.06 & 35.33 \\
& $\rho i$ & 0.33 & 0.10 & 14.21 \\
\hline
& $k_c$ & 0.46 & 0.01 & 62.29 \\
$\beta_{i,\perp}\sim1$ & $d_i$ & 0.47 & 0.03 & 23.45 \\
& $\rho_i$ & 0.47 & 0.03 & 23.32 \\
\vspace{-0.38cm}
\enddata
\end{deluxetable}

We now extend our analysis to include an entire year of data from 2012 in the same way as Section \ref{sec:break}. In Figure 7 we show histograms in the same format as Figure 3 for $f_L$ against $f_h$ for all data and then separated into periods of slow and fast wind in panels (a-c), (d-f), and (g-i), respectively. In Figure 8 we show in a similar fashion to Figure 4, $f_L$ against $f_h$ for periods where $\beta_{i,\perp}\sim1$. Finally, in Figure 9, we plot histograms for $f_L$ against $f_p$. Here, we discard data with $f_h\geq f_{noise}$ and $f_p\geq f_{noise}$ to ensure that instrumental noise does not affect our results, and data where $f_h\leq0.2$ and $f_p\leq0.2$ Hz, as discussed previously. We provide the results from our statistical analysis for $f_L$ and $f_h$ in Table \ref{tab:3} and for $f_L$ and $f_p$ in Table \ref{tab:4}.

We find that $f_{\rho i}$ has the lowest correlations and highest residuals with $f_h$ regardless of wind speed, which is consistent with Figure 7, where the distribution of data deviates significantly from the black dashed-line. We conclude that $f_{\rho i}$ is not directly comparable to $f_h$ within our studied interval. When we consider all data, $f_{kc}$ has the highest correlation coefficient of $R=0.48$ and lowest residual of $\rho=0.09$, compared to $R=0.40$ and $\rho=0.15$ for $f_{di}$. We find that 64.29\% of the total number of spectra fall within the two red dashed-lines for $f_{kc}$ in Figure 7(a), compared to 33.37\% for $f_{di}$ in panel (b). These percentages are similar regardless of wind speed. Besides similar correlation coefficients of about $R=0.42-0.45$ in fast wind streams, $f_{kc}$ is closer to the relationship $f_L\simeq f_h$ than $f_{di}$, from visual comparison of panels (g) and (h). In particular, we see that $f_{kc}$ has the lowest residual of $\rho=0.04$ during fast wind streams, compared to $\rho=0.06$ in slow wind. Comparing panels (d) and (g), the percentages of spectra within the red dashed-lines for $f_{kc}$ is lower in the fast wind at 61.38\% than in the slow wind where it is 64.72\%, despite a lower residual in the former.

As in the previous section, we also filter the data to include only periods where $0.95\geq\beta_{i,\perp}\leq1.05$ and show the corresponding 2D histograms in Figure 8, as well as the results from our statistical analysis in the bottom panel of Table \ref{tab:3}. Our results are similar to those for $f_b$ and $f_L$, where we see clearly that $f_{kc}$ best corresponds to $f_h$. From our statistical analysis, the correlation coefficients are the same for both $f_{di}$ and $f_{\rho i}$ at $R=0.47$, and almost the same at $R=0.46$ for $f_{kc}$, however, the latter again has the lowest residual at $\rho=0.01$, compared to $\rho=0.03$ for the other two scales. Again, we find a high percentage of the number of spectra where $f_b\simeq f_{kc}$ at 62.29\%, almost triple that of the other two scales.

Following Section \ref{sec:break}, we conclude that the onset of the helicity signature is also related to the cyclotron resonant scale and therefore, both the spectral steepening and coherent helicity signature are likely linked to the same physical process: proton-cyclotron resonance. This signature is most prevalent when the spacecraft measures fast wind streams, and therefore, we conclude that there is a stronger relationship between $f_{kc}$ and $f_h$ during these periods, as we also see for $f_{kc}$ and $f_b$. The lower percentage for $f_{kc}$ in fast wind is likely due to the reduced number of available measurements than for slow wind periods, as we see in plots in the right column of Figure 7, or because of the limited applicability of wind speed as the only criterion to categorize wind streams.

\begin{figure*}
\centering
\includegraphics[scale=0.8]{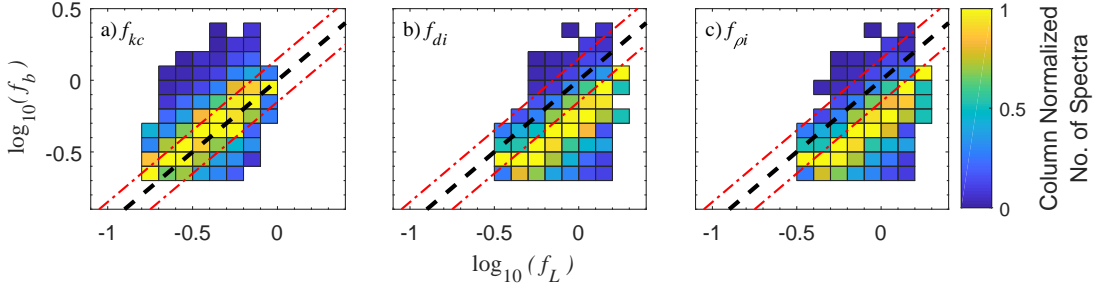}
\caption{Histograms for 2012 of the estimated helicity onset frequency, $f_h$, versus the three characteristic plasma scales, converted into frequencies using Taylor's hypothesis - $f_L$ represents $f_{kc}$, $f_{di}$ and $f_{\rho i}$, for each column respectively. The data used are for periods where $0.95\geq\beta_{i,\perp}\leq1.05$. The color-bar represents the column-normalized number of spectra. The black dashed lines represent $f_b=f_L$ and similarly, the red dashed lines are $f_b=f_L\;\sqrt[]{2}$ and $f_b=f_L/\,\sqrt[]{2}$, which give the resolution of the wavelet transform about the line $f_b=f_L$ due to the finite width of the Morlet wavelet in frequency space.}
\label{fig:8}
\end{figure*}

\begin{deluxetable}{lcccc}
\tablecaption{Correlation coefficients, residuals, and percentages for $f_L$ and $f_p$ from the data shown in Figure 9.\label{tab:4}}
\tablehead{
\colhead{} &
\colhead{Plasma Scale} &
\colhead{Correlation} &
\colhead{Residual} &
\colhead{Percent} \\
\colhead{} &
\colhead{$L$} &
\colhead{$R$} &
\colhead{$\rho$} &
\colhead{\%}
}
\startdata
& $k_c$ & 0.55 & 0.19 & 16.83 \\
All Data & $d_i$ & 0.51 & 0.12 & 44.79 \\
& $\rho_i$ & 0.34 & 0.15 & 46.62 \\
\hline
& $k_c$ & 0.56 & 0.11 & 19.52 \\
Slow & $d_i$ & 0.49 & 0.07 & 46.00 \\
& $\rho_i$ & 0.41 & 0.11 & 39.82 \\
\hline
& $k_c$ & 0.52 & 0.08 & 14.15 \\
Fast & $d_i$ & 0.53 & 0.05 & 41.86 \\
& $\rho i$ & 0.28 & 0.06 & 51.74 \\
\vspace{-0.38cm}
\enddata
\end{deluxetable}

Moving now to the peak frequency of the helicity signature and our results presented in Figure 9 and Table \ref{tab:4}, we find that both $f_{kc}$ and $f_{di}$ have similar correlation coefficients with $f_p$ of $R=0.49-0.56$, regardless of wind speed. We can also see little difference when comparing visually the three columns in the Figure for these two scales. However, the lowest residuals are seen for $f_{di}$, giving $\rho=0.07$ and $\rho=0.05$ during slow and fast wind streams, respectively. We find that $f_{\rho i}$ has the lowest correlation coefficients at $R=0.28-0.41$ compared to $f_{kc}$ and $f_{di}$. However, its residuals are comparable to that of $f_{di}$, at $\rho=0.06$ for fast wind and $\rho=0.11$ for slow wind. Figure 9 shows that $f_p$ correlates with both $f_{di}$ and $f_{\rho i}$, as expected since they differ only by a factor of $\sqrt[]{\beta_{i,\perp}}$, but there is a constant offset in frequency for $f_{di}$ that is not present for $f_{\rho i}$. When we consider all data, 46.62\% of spectra satisfy $f_p\simeq f_{\rho i}$ compared to 44.79\% for $f_p\simeq f_{di}$, within the \textit{e}-folding frequency.

For frequencies $f_{\rho i}>1$ Hz, the most likely value for $f_p$ diverges from the black dashed line in panel (c) of Figure 9, which results in the low values for $R$ with $f_{\rho i}$, and higher correlation with $f_{di}$. However, we find that 39.82\% of spectra in slow wind and 51.74\% in fast wind satisfy $f_p\simeq f_{\rho i}$ within the \textit{e}-folding frequency. The higher percentage in fast wind is likely due to the stronger helicity signature compared to slow wind, making detection easier. The divergence at high frequencies may be caused by under-sampling, because $f_p$ can exceed $f_{noise}$ at these frequencies, but we try to account for this by discarding bins with $\leq$10 spectra. We also see a similar feature in panel (i) of Figure 7 as in panel (i) in Figure 3. Due to the noise-floor, it is difficult to distinguish whether this feature is physical or an artifact. Despite this divergence at high frequencies, we conclude that $\rho_i$ best corresponds with the peak in the helicity signature at ion-kinetic frequencies compared to the other two scales.

\begin{figure*}
\centering
\includegraphics[scale=0.8]{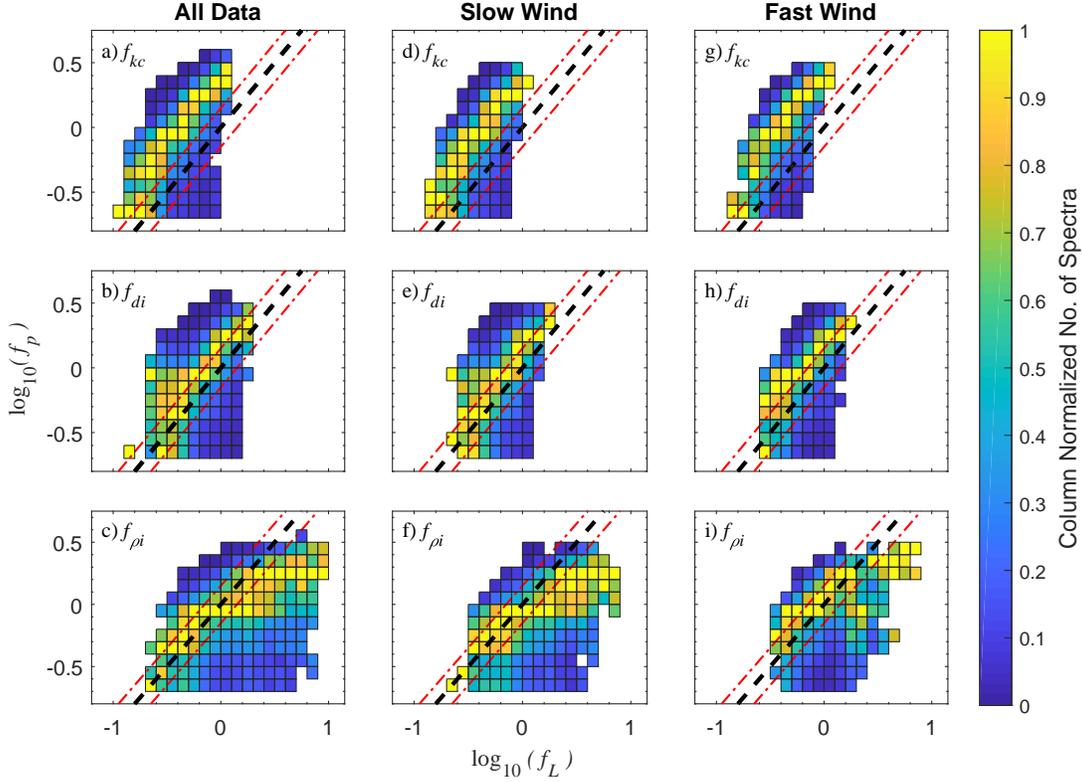}
\caption{(a-c) Histograms for 2012 of the estimated helicity peak frequency, $f_p$, versus the three characteristic plasma scales, converted into frequencies using Taylor's hypothesis - $f_L$ represents $f_{kc}$, $f_{di}$ and $f_{\rho i}$, for each row respectively. (d-f) The corresponding results for only slow wind (<400 km/s) periods. (g-i) The corresponding results for only fast wind (>500 km/s) streams. The color-bar represents the column-normalized number of spectra. The black dashed lines represent $f_p=f_L$ and similarly, the red dashed lines are $f_b=f_L\;\sqrt[]{2}$ and $f_b=f_L/\,\sqrt[]{2}$, which give the resolution of the wavelet transform about the line $f_b=f_L$.}
\label{fig:9}
\end{figure*}

\section{Discussion}

Our main result is the correlation of the cyclotron resonance scale, $1/k_c$, with the onset of the spectral steepening of the magnetic field fluctuation power spectrum and a coherent magnetic helicity signature at ion-kinetic scales. The helicity also reaches a maximum at scales comparable to $\rho_i$. Therefore, we suggest that these two signatures are related and result from proton-cyclotron damping of AICs, leading to a steeper power law due to dissipation at these scales. We then explain the resulting helicity signature as due to the residual population of KAWs left behind after the AICs are removed from the turbulent cascade. This cyclotron-resonant dissipation is consistent with the shape of proton distributions observed in the fast wind \citep[e.g.,][]{Tu2001OnCorona,Marsch2001,Tu2002,Marsch2004,Heuer2007DiffusionProtons,He2015,He2015a}. These results hold over most solar wind conditions, but in particular during periods of fast wind streams where the helicity signature is strongest.

We find that over the course of 2012, the onset of the coherent helicity signature corresponds to $1/k_c$ for 64.29\% of the time, within the limits of the \textit{e}-folding frequency of our Morlet wavelet. This value does not change significantly when we filter data according to wind speed. Given how we measure $f_h$ using a Gaussian function, there is no reason for both $f_h$ and $f_{kc}$ to correlate well by random chance. The onset of the helicity peak determined from the FWHM does not necessarily need to occur at $1/k_c$ and yet, we find they are both closely related. Similarly, we find that 52.08\% of the time the break scale corresponds to $1/k_c$. These results imply that cyclotron resonance with protons likely occurs at least half the time in the solar wind at ion-kinetic scales. However, the lower percentage for $f_b\simeq f_{kc}$ indicates that resonance with AICs may not always lead to a sufficient amount of energy to be removed from the cascade to result in a spectral steepening at $1/k_c$. Alternatively, it may be due to the higher level of uncertainty in measuring the break scale compared to the onset of the helicity signature. We can also interpret the square of correlation coefficient values, $R^2$, as a percentage of the time that a parameter depends on another. From our results, this would give somewhat lower percentages (23-31\%) than our first estimates. We do not place too much weight on these correlation coefficients since they can be misleading, especially when considering our results for periods where $\beta_{i,\perp}\sim1$. Therefore, we take our first calculation as a more reliable estimate.

The better agreement found in fast wind between $1/k_c$, the break scale, and the onset of the helicity signature suggests that cyclotron damping primarily occurs in fast wind streams, which are typically more Alfv\'enic with a higher population of AICs \citep{Roberts2015,Telloni2016,Lion2016a}. However, we find that the coherent helicity signature disappears or significantly weakens during slow wind periods, in agreement with \citet{Bruno2015}. The dispersion relation for Alfv\'en waves splits into KAWs or AICs at a critical angle to the magnetic field that is dependent on $\beta$ \citep{Gary1986Low-frequencyHelicity}. Therefore, an explanation for the reduction in the prevalence of helicity signatures in the slow wind may be due to different $\beta$ in fast and slow wind, affecting how we observe the helicity signature resulting from KAWs. Despite the lack of the coherent helicity signature in the slow wind, we still observe a spectral steepening at $1/k_c$, but the agreement is weaker than in the fast wind.

The anisotropic nature of plasma turbulence in the solar wind implies a limited role of $k_\parallel$ in the energy cascade in the inertial range, due to a higher amount of power in perpendicular wavenumbers \citep{Horbury2008,Chen2010a,Chen2010,Wicks2010}. Despite this, we find that during the interval of data we study, the break most often occurs at $k\simeq k_c$ and not $kd_i\simeq 1$ or $k\rho_i\simeq1$, as clearly shown during periods where $\beta_{i,\perp}\sim1$. This result is consistent with studies of turbulence at extreme $\beta_{i,\perp}$ \citep[e.g.,][]{Smith2001,Chen2014}. In their study, \citet{Chen2014} rule out the role of $k_c$ since they assume $k_\perp\gg k_\parallel$ at ion-kinetic scales. However, other studies show that the $k_\parallel$ component of the turbulence, while small compared to $k_\perp$, increases around ion-kinetic scales \citep{Bieber1996,Leamon1998a,Dasso2005AnisotropyFluctuations,Hamilton2008,Roberts2015}. Our results suggest that this small $k_\parallel$ component of the turbulence is damped from the cascade, which leads to the observed spectral steepening at these scales.

We note that previous studies \citep[e.g.,][]{Markovskii2008,Bourouaine2012,Bruno2014,Chen2014} include an additional $\sin{\theta_{Bv}}$ factor in the definition of the associated break scales in order to account for the anisotropic nature of the turbulence at kinetic scales, where $\theta_{Bv}$ is the angle between the magnetic field, $\bf{B}$, and velocity flow, $\textbf{v}_{sw}$. The inclusion of this factor in our analysis only slightly improves our correlations and lowers our residuals by about 10\% for $f_{di}$ and $f_{\rho i}$. Therefore, it is not necessary to include this factor since the agreement of $f_{kc}$ with $f_b$ or $f_h$ is still clearly better than the agreement of $f_{di}\sin{\theta_{Bv}}$ and $f_{\rho_i}\sin{\theta_{Bv}}$, showing that we cannot rule out cyclotron damping because of the anisotropy of inertial range turbulence.

Recent studies by \citet{Markovskii2015}, \citet{Markovskii2016MagneticTemperature}, and \citet{Markovskii2016TheWind} attribute the coherent helicity signature to two competing processes, one which generates and the other which destroys magnetic helicity: the generation of helicity is due to the increased compressional component of KAW fluctuations at small scales and development of a magnetic field component parallel to the local mean field \citep{Howes2010,TenBarge2012EvidenceSimulations,Markovskii2013MagneticTurbulence,Markovskii2013MagneticTurbulenceb}, while the decrease in helicity arises from the demagnetization of the protons from the magnetic field \citep{Vasquez2012VelocityRegime}. \citet{Markovskii2015} interpret the peak in the helicity as arising from the balance of these two processes. In a later study, \citet{Markovskii2016TheWind} find that this peak is best correlated with the gyroscale modified by the electron beta, $\beta_e\,$: $\rho_i=d_i/\,\sqrt[]{\beta_i+\beta_e}\,$, and is therefore affected by the total plasma pressure. We do not use electron data here, however, if $\beta_i=\beta_e$, then $\beta_e$ will contribute a maximum factor of $1/\,\sqrt[]{2}$, roughly equivalent to the uncertainty in our results from the use of a Morlet wavelet. We see that the coherent helicity signature disappears towards smaller scales than $\rho_i$.Therefore, there is no significant difference in our results and overall accuracy by excluding electron data.

We cannot rule out that a combination of processes may lead to the observed helicity signature, for example, from the increased compressional component of KAWs \citep[e.g.,][]{Howes2010}, or from the presence of magnetosonic/whistler waves \citep{Podesta2011a,Podesta2011}. However, our results suggest the dominant cause for the onset of the observed signature is due to proton-cyclotron resonance with AICs. We do not investigate the origin of these cyclotron-resonant fluctuations, but rather show evidence for their existence and subsequent dissipation. We also see that the coherent helicity signature disappears towards smaller scales than $\rho_i$. The disappearance of the signature at higher frequencies may be due to the demagnetization of protons \citep{Vasquez2012VelocityRegime}, the increasing balance of sunward and anti-sunward energy fluxes at smaller scales \citep{He2012}, aliasing of power \citep{Russell1972,Klein2014}, or instrumental noise. We are unable to determine the cause of the return of the helicity to around zero from this study.

Our results also indicate that the transition range following the spectral break in the magnetic field power spectrum often seen in fast wind streams is due to proton-cyclotron resonance. This link between the transition range and cyclotron resonance is consistent with the findings of several other studies \citep{Podesta2009,Bourouaine2010CorrelationsWind,Bruno2014a,Bruno2015,Roberts2017DirectTurbulence}. We note that our results do not rule out the role of other non-linear wave-particle interactions and kinetic non-resonant mechanisms causing dissipation or dispersive effects. In fact, past studies by \citet{Leamon1998,Leamon1998a,Leamon1999,Leamon2000,Smith2012} have shown non-resonant damping (e.g., Landau or transit-time damping) of ions and electrons likely accounts for the remaining $\sim$50\% of dissipation, which is consistent with our findings that 52.13\% of the time cyclotron-resonant damping is occurring.

\section{Summary and Conclusions}

We use magnetic field and particle moment data from the MFI and SWE instruments onboard the \textit{Wind} spacecraft to study the nature of the solar wind turbulence at ion-kinetic scales. We first analyze solar wind data from 2012, investigating the spectral properties of the magnetic field. We use a Morlet continuous wavelet transform to compute the power and normalized magnetic helicity spectra for successive 92 s intervals. To determine whether spectral features are physical at high frequencies, we identify the noise-floor of the MFI instrument using tail-lobe crossings of the Earth's magnetosphere from early 2004, finding it at a higher amplitude than originally predicted. Finally, we use particle data at the same 92 s cadence to calculate the characteristic proton scales, $1/k_c$, $d_i$, and $\rho_i$, and investigate their relationship with the spectral break and coherent helicity signature at ion-kinetic scales.

The automated routine we use to analyze solar wind magnetic power and helicity spectra combines both the identification of the break scale and analysis of the properties of the magnetic fluctuations at ion-kinetic scales. This analysis of high-resolution spectra accounts for the variability of the plasma scales under different solar wind conditions, while also processing large volumes of solar wind data. For the first time, we link the spectral break frequency, helicity onset, and cyclotron resonance scale. We expand on past results by investigating both fast and slow wind streams, as well as periods where $\beta_{i,\perp}\sim$1.

In agreement with \citet{Bruno2014,Bruno2015,Telloni2015}, we find that the high-frequency spectral steepening in a fast wind stream is best associated with the cyclotron resonance scale, $1/k_c$, which also forms the low-frequency bound of a coherent helicity signature. We show for the first time, these results hold in general for fast streams and to a lesser extent, for slow wind, where the helicity signature weakens or disappears completely. We also find that the peak of the helicity enhancement is associated with the ion gyroscale, $\rho_i$, consistent with the findings of \citet{Markovskii2015}, again best seen within fast streams, where the enhancement in the helicity is strongest.

Our key result presented here is evidence supporting proton-cyclotron resonant damping as a dissipation mechanism of solar wind turbulence at ion-kinetic scales, occurring at least half the time in the solar wind. This resonance results in the damping of Alfv\'en/ion-cyclotron waves, particularly in the more Alfv\'enic fast wind, leading to the steepening of the magnetic field fluctuation power spectrum. Therefore, we suggest that the AICs are removed from the turbulence at these scales, resulting in a coherent helicity spectrum from the remaining KAWs, which are not cyclotron resonant. We note that we do not speculate about the origin of the cyclotron-resonant fluctuations, but rather show evidence for their existence and dissipation.

Further investigative work is on-going to determine the relative importance of proton-cyclotron resonance for the dissipation of turbulence and subsequent heating of the particle distributions. In particular, we still need to quantify the energy dissipated and the amount of energy that continues to cascade down to electron scales. We leave this work to a subsequent study. Understanding the nature of dissipation of the turbulence in the solar wind will provide us with a deeper understanding of the macroscopic properties of the solar wind and insight into similar processes in other collisionless plasmas. The future Solar Orbiter and Parker Solar Probe missions will also help us to explore these important areas of heliophysics research.

\acknowledgments
L.D.W. is funded by an STFC Studentship; D.V. is supported by an STFC Ernest Rutherford Fellowship, ST/P003826/1; C.J.O. is supported by the STFC consolidated grant to UCL/MSSL, ST/N000722/1. The authors thank R. L. Alexander for providing suitable intervals for our noise-floor analysis and the MFI/SWE instrument teams for provision of the data. The authors also acknowledge A. R. Macneil, G. A. Graham, N. M. E. Kalmoni, O. W. Roberts, C. H. K. Chen, and H. Wu for useful comments and discussions. Data from the WIND spacecraft were obtained from the \href{http://spdf.gsfc.nasa.gov}{SPDF web-site}.

\bibliographystyle{yahapj}
\bibliography{Mendeley}

\begin{thebibliography}{}
\providecommand\natexlab[1]{#1}
\providecommand\JournalTitle[1]{#1}

\bibitem[{Acu{\~{n}}a {et~al.}(1995)Acu{\~{n}}a, Ogilvie, Baker, Curtis,
  Fairfield, \& Mish}]{Acuna1995}
Acu{\~{n}}a, M.~H., Ogilvie, K.~W., Baker, D.~N., {et~al.} 1995,
  \href{http://dx.doi.org/10.1007/BF00751323}{\JournalTitle{Space Science
  Reviews}, 71, 5}

\bibitem[{Alexandrova {et~al.}(2013)Alexandrova, Chen, Sorriso-Valvo, Horbury,
  \& Bale}]{Alexandrova2013}
Alexandrova, O., Chen, C. H.~K., Sorriso-Valvo, L., Horbury, T.~S., \& Bale,
  S.~D. 2013,
  \href{http://dx.doi.org/10.1007/s11214-013-0004-8}{\JournalTitle{Space
  Science Reviews}, 178, 101}

\bibitem[{Bale {et~al.}(2009)Bale, Kasper, Howes, Quataert, Salem, \&
  Sundkvist}]{Bale2009}
Bale, S.~D., Kasper, J.~C., Howes, G.~G., {et~al.} 2009,
  \href{http://dx.doi.org/10.1103/PhysRevLett.103.211101}{\JournalTitle{Physical
  Review Letters}, 103}

\bibitem[{Batchelor(1970)}]{Batchelor1970}
Batchelor, G.~K. 1970, {The Theory of Homogenous Turbulence} (Cambridge
  University Press)

\bibitem[{Bennett(1948)}]{Bennett1948SpectraSignals}
Bennett, W.~R. 1948,
  \href{http://dx.doi.org/10.1002/j.1538-7305.1948.tb01340.x}{\JournalTitle{Bell
  System Technical Journal}, 27, 446}

\bibitem[{Bieber {et~al.}(1996)Bieber, Wanner, \& Matthaeus}]{Bieber1996}
Bieber, J.~W., Wanner, W., \& Matthaeus, W.~H. 1996,
  \href{http://dx.doi.org/10.1029/95JA02588}{\JournalTitle{Journal of
  Geophysical Research}, 101, 2511}

\bibitem[{Boldyrev {et~al.}(2013)Boldyrev, Horaites, Xia, \&
  Perez}]{Boldyrev2013}
Boldyrev, S., Horaites, K., Xia, Q., \& Perez, J.~C. 2013,
  \href{http://dx.doi.org/10.1088/0004-637X/777/1/41}{\JournalTitle{The
  Astrophysical Journal}, 777, 41}

\bibitem[{Boldyrev \& Perez(2012)}]{Boldyrev2012}
Boldyrev, S., \& Perez, J.~C. 2012,
  \href{http://dx.doi.org/10.1088/2041-8205/758/2/L44}{\JournalTitle{The
  Astrophysical Journal}, 758, L44}

\bibitem[{Bourouaine {et~al.}(2012)Bourouaine, Alexandrova, Marsch, \&
  Maksimovic}]{Bourouaine2012}
Bourouaine, S., Alexandrova, O., Marsch, E., \& Maksimovic, M. 2012,
  \href{http://dx.doi.org/10.1088/0004-637X/749/2/102}{\JournalTitle{The
  Astrophysical Journal}, 749, 102}

\bibitem[{Bourouaine {et~al.}(2010)Bourouaine, Marsch, \&
  Neubauer}]{Bourouaine2010CorrelationsWind}
Bourouaine, S., Marsch, E., \& Neubauer, F.~M. 2010,
  \href{http://dx.doi.org/10.1029/2010GL043697}{\JournalTitle{Geophysical
  Research Letters}, 37, 1}

\bibitem[{Brandenburg {et~al.}(2011)Brandenburg, Subramanian, Balogh, \&
  Goldstein}]{Brandenburg2011ScaleWind}
Brandenburg, A., Subramanian, K., Balogh, A., \& Goldstein, M.~L. 2011,
  \href{http://dx.doi.org/10.1088/0004-637X/734/1/9}{\JournalTitle{The
  Astrophysical Journal}, 734, 9}

\bibitem[{Bruno \& Carbone(2013)}]{Bruno2013}
Bruno, R., \& Carbone, V. 2013,
  \href{http://dx.doi.org/10.12942/lrsp-2013-2}{\JournalTitle{Living Reviews in
  Solar Physics}, 10}

\bibitem[{Bruno \& Telloni(2015)}]{Bruno2015}
Bruno, R., \& Telloni, D. 2015,
  \href{http://dx.doi.org/10.1088/2041-8205/811/2/L17}{\JournalTitle{The
  Astrophysical Journal Letters}, 811, L17}

\bibitem[{Bruno {et~al.}(2017)Bruno, Telloni, DeIure, \&
  Pietropaolo}]{Bruno2017SolarScales}
Bruno, R., Telloni, D., DeIure, D., \& Pietropaolo, E. 2017,
  \href{http://dx.doi.org/10.1093/mnras/stx2008}{\JournalTitle{Monthly Notices
  of the Royal Astronomical Society}, 472, 1052}

\bibitem[{Bruno \& Trenchi(2014)}]{Bruno2014}
Bruno, R., \& Trenchi, L. 2014,
  \href{http://dx.doi.org/10.1088/2041-8205/787/2/L24}{\JournalTitle{The
  Astrophysical Journal Letters}, 787, L24}

\bibitem[{Bruno {et~al.}(2014)Bruno, Trenchi, \& Telloni}]{Bruno2014a}
Bruno, R., Trenchi, L., \& Telloni, D. 2014,
  \href{http://dx.doi.org/10.1088/2041-8205/793/1/L15}{\JournalTitle{The
  Astrophysical Journal Letters}, 793, L15}

\bibitem[{Cerri {et~al.}(2016)Cerri, Califano, Jenko, Told, \&
  Rincon}]{Cerri2016Subproton-ScaleSimulations}
Cerri, S.~S., Califano, F., Jenko, F., Told, D., \& Rincon, F. 2016,
  \href{http://dx.doi.org/10.3847/2041-8205/822/1/L12}{\JournalTitle{The
  Astrophysical Journal Letters}, 822, L12}

\bibitem[{Chandran {et~al.}(2010)Chandran, Li, Rogers, Quataert, \&
  Germaschewski}]{Chandran2010}
Chandran, B. D.~G., Li, B., Rogers, B.~N., Quataert, E., \& Germaschewski, K.
  2010, \href{http://dx.doi.org/10.1088/0004-637X/720/1/503}{\JournalTitle{The
  Astrophysical Journal}, 720, 503}

\bibitem[{Chen {et~al.}(2010{\natexlab{a}})Chen, Horbury, Schekochihin, Wicks,
  Alexandrova, \& Mitchell}]{Chen2010a}
Chen, C. H.~K., Horbury, T.~S., Schekochihin, A.~A., {et~al.}
  2010{\natexlab{a}},
  \href{http://dx.doi.org/10.1103/PhysRevLett.104.255002}{\JournalTitle{Physical
  Review Letters}, 104, 255002}

\bibitem[{Chen {et~al.}(2014)Chen, Leung, Boldyrev, Maruca, \& Bale}]{Chen2014}
Chen, C. H.~K., Leung, L., Boldyrev, S., Maruca, B.~A., \& Bale, S.~D. 2014,
  \href{http://dx.doi.org/10.1002/2014GL062009}{\JournalTitle{Geophysical
  Research Letters}, 41, 8081}

\bibitem[{Chen {et~al.}(2016)Chen, Matteini, Schekochihin, Stevens, Salem,
  Maruca, Kunz, \& Bale}]{Chen2016}
Chen, C. H.~K., Matteini, L., Schekochihin, A.~A., {et~al.} 2016,
  \href{http://dx.doi.org/10.3847/2041-8205/825/2/L26}{\JournalTitle{The
  Astrophysical Journal Letters}, 825, L26}

\bibitem[{Chen {et~al.}(2010{\natexlab{b}})Chen, Wicks, Horbury, \&
  Schekochihin}]{Chen2010}
Chen, C. H.~K., Wicks, R.~T., Horbury, T.~S., \& Schekochihin, A.~A.
  2010{\natexlab{b}},
  \href{http://dx.doi.org/10.1088/2041-8205/711/2/L79}{\JournalTitle{The
  Astrophysical Journal}, 711, L79}

\bibitem[{Coleman(1968)}]{ColemanJr.1968}
Coleman, P.~J. 1968, \href{http://dx.doi.org/10.1086/149674}{\JournalTitle{The
  Astrophysical Journal}, 153, 371}

\bibitem[{Cranmer {et~al.}(2009)Cranmer, Matthaeus, Breech, \&
  Kasper}]{Cranmer2009}
Cranmer, S.~R., Matthaeus, W.~H., Breech, B.~A., \& Kasper, J.~C. 2009,
  \href{http://dx.doi.org/10.1088/0004-637X/702/2/1604}{\JournalTitle{The
  Astrophysical Journal}, 702, 1604}

\bibitem[{Dasso {et~al.}(2005)Dasso, Milano, Matthaeus, \&
  Smith}]{Dasso2005AnisotropyFluctuations}
Dasso, S., Milano, L.~J., Matthaeus, W.~H., \& Smith, C.~W. 2005,
  \href{http://dx.doi.org/10.1086/499559}{\JournalTitle{The Astrophysical
  Journal}, 635, L181}

\bibitem[{Denskat {et~al.}(1983)Denskat, Beinroth, \& Neubauer}]{Denskat1983}
Denskat, K.~U., Beinroth, H.~J., \& Neubauer, F.~M. 1983, \JournalTitle{Journal
  of Geophysics}, 54, 60

\bibitem[{Dmitruk {et~al.}(2004)Dmitruk, Matthaeus, \& Seenu}]{Dmitruk2004}
Dmitruk, P., Matthaeus, W.~H., \& Seenu, N. 2004,
  \href{http://dx.doi.org/10.1086/425301}{\JournalTitle{The Astrophysical
  Journal}, 617, 667}

\bibitem[{Forman {et~al.}(2011)Forman, Wicks, \& Horbury}]{Forman2011}
Forman, M.~A., Wicks, R.~T., \& Horbury, T.~S. 2011,
  \href{http://dx.doi.org/10.1088/0004-637X/733/2/76}{\JournalTitle{The
  Astrophysical Journal}, 733, 76}

\bibitem[{Franci {et~al.}(2016)Franci, Landi, Matteini, Verdini, \&
  Hellinger}]{Franci2016PlasmaSimulations}
Franci, L., Landi, S., Matteini, L., Verdini, A., \& Hellinger, P. 2016,
  \href{http://dx.doi.org/10.3847/1538-4357/833/1/91}{\JournalTitle{The
  Astrophysical Journal}, 833, 91}

\bibitem[{Franci {et~al.}(2017)Franci, Cerri, Califano, Landi, Papini, Verdini,
  Matteini, Jenko, \& Hellinger}]{Franci2017MagneticTurbulence}
Franci, L., Cerri, S.~S., Califano, F., {et~al.} 2017,
  \href{http://dx.doi.org/10.3847/2041-8213/aa93fb}{\JournalTitle{The
  Astrophysical Journal Letters}, 850, L16}

\bibitem[{Galtier(2006)}]{Galtier2006}
Galtier, S. 2006,
  \href{http://dx.doi.org/10.1017/S0022377806004521}{\JournalTitle{Journal of
  Plasma Physics}, 72, 721}

\bibitem[{Galtier \& Buchlin(2007)}]{Galtier2007}
Galtier, S., \& Buchlin, E. 2007,
  \href{http://dx.doi.org/10.1086/510423}{\JournalTitle{The Astrophysical
  Journal}, 656, 560}

\bibitem[{Gary(1986)}]{Gary1986Low-frequencyHelicity}
Gary, S.~P. 1986,
  \href{http://dx.doi.org/10.1017/S0022377800011442}{\JournalTitle{Journal of
  Plasma Physics}, 35, 431}

\bibitem[{Gary(1993)}]{Gary1993}
---. 1993, {Theory of Space Plasma Microinstabilities} (Cambridge University
  Press)

\bibitem[{Gary(1999)}]{Gary1999CollisionlessTheory}
---. 1999, \href{http://dx.doi.org/10.1029/1998JA900161}{\JournalTitle{Journal
  of Geophysical Research}, 104, 6759}

\bibitem[{Gary(2015)}]{Gary2015}
---. 2015,
  \href{http://dx.doi.org/10.1098/rsta.2014.0149}{\JournalTitle{Philosophical
  Transactions of the Royal Society A: Mathematical, Physical and Engineering
  Sciences}, 373, 20140149}

\bibitem[{Gary \& Borovsky(2004)}]{Gary2004}
Gary, S.~P., \& Borovsky, J.~E. 2004,
  \href{http://dx.doi.org/10.1029/2004JA010399}{\JournalTitle{Journal of
  Geophysical Research: Space Physics}, 109, A06105}

\bibitem[{Gary \& Smith(2009)}]{Gary2009}
Gary, S.~P., \& Smith, C.~W. 2009,
  \href{http://dx.doi.org/10.1029/2009JA014525}{\JournalTitle{Journal of
  Geophysical Research: Space Physics}, 114, 1}

\bibitem[{Ghosh {et~al.}(1996)Ghosh, Siregar, Roberts, \&
  Goldstein}]{Ghosh1996SimulationMagnetohydrodynamics}
Ghosh, S., Siregar, E., Roberts, D.~A., \& Goldstein, M.~L. 1996,
  \href{http://dx.doi.org/10.1029/95JA03201}{\JournalTitle{Journal of
  Geophysical Research}, 101, 2493}

\bibitem[{Goldstein {et~al.}(1994)Goldstein, Roberts, \& Fitch}]{Goldstein1994}
Goldstein, M.~L., Roberts, D.~A., \& Fitch, C.~A. 1994,
  \href{http://dx.doi.org/10.1029/94JA00789}{\JournalTitle{Journal of
  Geophysical Research}, 99, 519}

\bibitem[{Goldstein {et~al.}(1995)Goldstein, Roberts, \&
  Matthaeus}]{Goldstein1995}
Goldstein, M.~L., Roberts, D.~A., \& Matthaeus, W.~H. 1995,
  \href{http://dx.doi.org/10.1146/annurev.aa.33.090195.001435}{\JournalTitle{Annual
  Review of Astronomy and Astrophysics}, 33, 283}

\bibitem[{Habbal {et~al.}(1997)Habbal, Woo, Fineschi, O'Neal, Kohl, Noci, \&
  Korendyke}]{Habbal1997OriginsWind}
Habbal, S.~R., Woo, R., Fineschi, S., {et~al.} 1997,
  \href{http://dx.doi.org/10.1086/310970}{\JournalTitle{The Astrophysical
  Journal}, 489, L103}

\bibitem[{Hamilton {et~al.}(2008)Hamilton, Smith, Vasquez, \&
  Leamon}]{Hamilton2008}
Hamilton, K., Smith, C.~W., Vasquez, B.~J., \& Leamon, R.~J. 2008,
  \href{http://dx.doi.org/10.1029/2007JA012559}{\JournalTitle{Journal of
  Geophysical Research: Space Physics}, 113, A01106}

\bibitem[{He {et~al.}(2011)He, Marsch, Tu, Yao, \& Tian}]{He2011}
He, J., Marsch, E., Tu, C.~Y., Yao, S., \& Tian, H. 2011,
  \href{http://dx.doi.org/10.1088/0004-637X/731/2/85}{\JournalTitle{The
  Astrophysical Journal}, 731, 85}

\bibitem[{He {et~al.}(2012{\natexlab{a}})He, Tu, Marsch, \& Yao}]{He2012a}
He, J., Tu, C.~Y., Marsch, E., \& Yao, S. 2012{\natexlab{a}},
  \href{http://dx.doi.org/10.1088/2041-8205/745/1/L8}{\JournalTitle{The
  Astrophysical Journal Letters}, 8}

\bibitem[{He {et~al.}(2012{\natexlab{b}})He, Tu, Marsch, \& Yao}]{He2012}
---. 2012{\natexlab{b}},
  \href{http://dx.doi.org/10.1088/0004-637X/749/1/86}{\JournalTitle{The
  Astrophysical Journal}, 749, 86}

\bibitem[{He {et~al.}(2015{\natexlab{a}})He, Wang, Tu, Marsch, \&
  Zong}]{He2015}
He, J., Wang, L., Tu, C.~Y., Marsch, E., \& Zong, Q. 2015{\natexlab{a}},
  \href{http://dx.doi.org/10.1088/2041-8205/800/2/L31}{\JournalTitle{The
  Astrophysical Journal}, 800, L31}

\bibitem[{He {et~al.}(2015{\natexlab{b}})He, Tu, Marsch, Chen, Wang, Pei,
  Zhang, Salem, \& Bale}]{He2015a}
He, J., Tu, C.~Y., Marsch, E., {et~al.} 2015{\natexlab{b}},
  \href{http://dx.doi.org/10.1088/2041-8205/813/2/L30}{\JournalTitle{The
  Astrophysical Journal Letters}, 813, L30}

\bibitem[{Hellinger {et~al.}(2006)Hellinger, Tr{\'{a}}vn{\'{i}}{\v{c}}ek,
  Kasper, \& Lazarus}]{Hellinger2006}
Hellinger, P., Tr{\'{a}}vn{\'{i}}{\v{c}}ek, P.~M., Kasper, J.~C., \& Lazarus,
  A.~J. 2006,
  \href{http://dx.doi.org/10.1029/2006GL025925}{\JournalTitle{Geophysical
  Research Letters}, 33, L09101}

\bibitem[{Heuer \& Marsch(2007)}]{Heuer2007DiffusionProtons}
Heuer, M., \& Marsch, E. 2007,
  \href{http://dx.doi.org/10.1029/2006JA011979}{\JournalTitle{Journal of
  Geophysical Research}, 112, 1}

\bibitem[{Horbury {et~al.}(2008)Horbury, Forman, \& Oughton}]{Horbury2008}
Horbury, T.~S., Forman, M.~A., \& Oughton, S. 2008,
  \href{http://dx.doi.org/10.1103/PhysRevLett.101.175005}{\JournalTitle{Physical
  Review Letters}, 101}

\bibitem[{Howes {et~al.}(2008{\natexlab{a}})Howes, Cowley, Dorland, Hammett,
  Quataert, \& Schekochihin}]{Howes2008}
Howes, G.~G., Cowley, S.~C., Dorland, W., {et~al.} 2008{\natexlab{a}},
  \href{http://dx.doi.org/10.1029/2007JA012665}{\JournalTitle{Journal of
  Geophysical Research: Space Physics}, 113}

\bibitem[{Howes {et~al.}(2008{\natexlab{b}})Howes, Dorland, Cowley, Hammett,
  Quataert, Schekochihin, \& Tatsuno}]{Howes2008KineticPlasmas}
Howes, G.~G., Dorland, W., Cowley, S.~C., {et~al.} 2008{\natexlab{b}},
  \href{http://dx.doi.org/10.1103/PhysRevLett.100.065004}{\JournalTitle{Physical
  Review Letters}, 100, 065004}

\bibitem[{Howes \& Quataert(2010)}]{Howes2010}
Howes, G.~G., \& Quataert, E. 2010,
  \href{http://dx.doi.org/10.1088/2041-8205/709/1/L49}{\JournalTitle{The
  Astrophysical Journal Letters}, 709, L49}

\bibitem[{Jian {et~al.}(2010)Jian, Russell, Luhmann, Anderson, Boardsen,
  Strangeway, Cowee, \& Wennmacher}]{Jian2010ObservationsAU}
Jian, L.~K., Russell, C.~T., Luhmann, J.~G., {et~al.} 2010,
  \href{http://dx.doi.org/10.1029/2010JA015737}{\JournalTitle{Journal of
  Geophysical Research: Space Physics}, 115, 1}

\bibitem[{Jian {et~al.}(2009)Jian, Russell, Luhmann, Strangeway, Leisner, \&
  Galvin}]{Jian2009}
---. 2009,
  \href{http://dx.doi.org/10.1088/0004-637X/701/2/L105}{\JournalTitle{The
  Astrophysical Journal}, 701, L105}

\bibitem[{Jian {et~al.}(2014)Jian, Wei, Russell, Luhmann, Klecker, Omidi,
  Isenberg, Goldstein, Figueroa-Vi{\~{n}}as, \& Blanco-Cano}]{Jian2014}
Jian, L.~K., Wei, H.~Y., Russell, C.~T., {et~al.} 2014,
  \href{http://dx.doi.org/10.1088/0004-637X/786/2/123}{\JournalTitle{The
  Astrophysical Journal}, 786, 123}

\bibitem[{Kasper {et~al.}(2002)Kasper, Lazarus, \& Gary}]{Kasper2002a}
Kasper, J.~C., Lazarus, A.~J., \& Gary, S.~P. 2002,
  \href{http://dx.doi.org/10.1029/2002GL015128}{\JournalTitle{Geophysical
  Research Letters}, 29, 3}

\bibitem[{Kiyani {et~al.}(2009)Kiyani, Chapman, Khotyaintsev, Dunlop, \&
  Sahraoui}]{Kiyani2009GlobalTurbulence}
Kiyani, K.~H., Chapman, S.~C., Khotyaintsev, Y.~V., Dunlop, M.~W., \& Sahraoui,
  F. 2009,
  \href{http://dx.doi.org/10.1103/PhysRevLett.103.075006}{\JournalTitle{Physical
  Review Letters}, 103, 075006}

\bibitem[{Kiyani {et~al.}(2013)Kiyani, Chapman, Sahraoui, Hnat, Fauvarque, \&
  Khotyaintsev}]{Kiyani2013EnhancedTurbulence}
Kiyani, K.~H., Chapman, S.~C., Sahraoui, F., {et~al.} 2013,
  \href{http://dx.doi.org/10.1088/0004-637X/763/1/10}{\JournalTitle{The
  Astrophysical Journal}, 763, 10}

\bibitem[{Kiyani {et~al.}(2015)Kiyani, Osman, \& Chapman}]{Kiyani2015}
Kiyani, K.~H., Osman, K.~T., \& Chapman, S.~C. 2015,
  \href{http://dx.doi.org/10.1098/rsta.2014.0155}{\JournalTitle{Philosophical
  Transactions of the Royal Society A: Mathematical, Physical and Engineering
  Sciences}, 373, 20140155}

\bibitem[{Klein {et~al.}(2014)Klein, Howes, TenBarge, \& Podesta}]{Klein2014}
Klein, K.~G., Howes, G.~G., TenBarge, J.~M., \& Podesta, J.~J. 2014,
  \href{http://dx.doi.org/10.1088/0004-637X/785/2/138}{\JournalTitle{The
  Astrophysical Journal}, 785, 138}

\bibitem[{Koval \& Szabo(2013)}]{Koval2013}
Koval, A., \& Szabo, A. 2013, \href{http://dx.doi.org/10.1063/1.4811025}{in AIP
  Conference Proceedings, Vol. 1539}, 211

\bibitem[{Leamon {et~al.}(1998{\natexlab{a}})Leamon, Matthaeus, Smith, \&
  Wong}]{Leamon1998}
Leamon, R.~J., Matthaeus, W.~H., Smith, C.~W., \& Wong, H.~K.
  1998{\natexlab{a}}, \href{http://dx.doi.org/10.1086/311698}{\JournalTitle{The
  Astrophysical Journal}, 181}

\bibitem[{Leamon {et~al.}(2000)Leamon, Matthaeus, Smith, Zank, Mullan, \&
  Oughton}]{Leamon2000}
Leamon, R.~J., Matthaeus, W.~H., Smith, C.~W., {et~al.} 2000,
  \href{http://dx.doi.org/10.1086/309059}{\JournalTitle{The Astrophysical
  Journal}, 537, 1054}

\bibitem[{Leamon {et~al.}(1998{\natexlab{b}})Leamon, Smith, Ness, Matthaeus, \&
  Wong}]{Leamon1998a}
Leamon, R.~J., Smith, C.~W., Ness, N.~F., Matthaeus, W.~H., \& Wong, H.~K.
  1998{\natexlab{b}},
  \href{http://dx.doi.org/10.1029/97JA03394}{\JournalTitle{Journal of
  Geophysical Research}, 103, 4775}

\bibitem[{Leamon {et~al.}(1999)Leamon, Smith, Ness, \& Wong}]{Leamon1999}
Leamon, R.~J., Smith, C.~W., Ness, N.~F., \& Wong, H.~K. 1999,
  \href{http://dx.doi.org/10.1029/1999JA900158}{\JournalTitle{Journal of
  Geophysical Research: Space Physics}, 104, 22331}

\bibitem[{Lepping {et~al.}(1995)Lepping, Acu{\~{n}}a, Burlaga, Farrell, Slavin,
  Schatten, Mariani, Ness, Neubauer, Whang, Byrnes, Kennon, Panetta, Scheifele,
  \& Worley}]{Lepping1995}
Lepping, R.~P., Acu{\~{n}}a, M.~H., Burlaga, L.~F., {et~al.} 1995,
  \href{http://dx.doi.org/10.1007/BF00751330}{\JournalTitle{Space Science
  Reviews}, 71, 207}

\bibitem[{Lion {et~al.}(2016)Lion, Alexandrova, \& Zaslavsky}]{Lion2016a}
Lion, S., Alexandrova, O., \& Zaslavsky, A. 2016,
  \href{http://dx.doi.org/10.3847/0004-637X/824/1/47}{\JournalTitle{The
  Astrophysical Journal}, 824, 47}

\bibitem[{MacBride {et~al.}(2008)MacBride, Smith, \& Forman}]{MacBride2008}
MacBride, B.~T., Smith, C.~W., \& Forman, M.~A. 2008,
  \href{http://dx.doi.org/10.1086/529575}{\JournalTitle{The Astrophysical
  Journal}, 679, 1644}

\bibitem[{Markovskii \&
  Vasquez(2013{\natexlab{a}})}]{Markovskii2013MagneticTurbulence}
Markovskii, S.~A., \& Vasquez, B.~J. 2013{\natexlab{a}},
  \href{http://dx.doi.org/10.1088/0004-637X/768/1/62}{\JournalTitle{The
  Astrophysical Journal}, 768, 62}

\bibitem[{Markovskii \&
  Vasquez(2013{\natexlab{b}})}]{Markovskii2013MagneticTurbulenceb}
Markovskii, S.~A., \& Vasquez, B.~J. 2013{\natexlab{b}},
  \href{http://dx.doi.org/10.1063/1.4811032}{in Solar Wind 13, Vol. 1539}, 239

\bibitem[{Markovskii \& Vasquez(2016)}]{Markovskii2016MagneticTemperature}
---. 2016,
  \href{http://dx.doi.org/10.3847/0004-637X/820/1/15}{\JournalTitle{The
  Astrophysical Journal}, 820, 15}

\bibitem[{Markovskii {et~al.}(2008)Markovskii, Vasquez, \&
  Smith}]{Markovskii2008}
Markovskii, S.~A., Vasquez, B.~J., \& Smith, C.~W. 2008,
  \href{http://dx.doi.org/10.1086/527431}{\JournalTitle{The Astrophysical
  Journal}, 675, 1576}

\bibitem[{Markovskii {et~al.}(2015)Markovskii, Vasquez, \&
  Smith}]{Markovskii2015}
---. 2015,
  \href{http://dx.doi.org/10.1088/0004-637X/806/1/78}{\JournalTitle{The
  Astrophysical Journal}, 806, 78}

\bibitem[{Markovskii {et~al.}(2016)Markovskii, Vasquez, \&
  Smith}]{Markovskii2016TheWind}
---. 2016,
  \href{http://dx.doi.org/10.3847/1538-4357/833/2/212}{\JournalTitle{The
  Astrophysical Journal}, 833, 212}

\bibitem[{Marsch(2006)}]{Marsch2006}
Marsch, E. 2006,
  \href{http://dx.doi.org/10.12942/lrsp-2006-1}{\JournalTitle{Living Reviews in
  Solar Physics}, 3, 1}

\bibitem[{Marsch(2012)}]{Marsch2012}
---. 2012,
  \href{http://dx.doi.org/10.1007/s11214-010-9734-z}{\JournalTitle{Space
  Science Reviews}, 172, 23}

\bibitem[{Marsch {et~al.}(2004)Marsch, Ao, \& Tu}]{Marsch2004}
Marsch, E., Ao, X.~Z., \& Tu, C.~Y. 2004,
  \href{http://dx.doi.org/10.1029/2003JA010330}{\JournalTitle{Journal of
  Geophysical Research}, 109, A04102}

\bibitem[{Marsch {et~al.}(1982{\natexlab{a}})Marsch, Goertz, \&
  Richter}]{Marsch1982}
Marsch, E., Goertz, C.~K., \& Richter, K. 1982{\natexlab{a}},
  \href{http://dx.doi.org/10.1029/JA087iA07p05030}{\JournalTitle{Journal of
  Geophysical Research}, 87, 5030}

\bibitem[{Marsch {et~al.}(1982{\natexlab{b}})Marsch, M{\"{u}}hlh{\"{a}}user,
  Schwenn, Rosenbauer, Pilipp, \& Neubauer}]{Marsch1982b}
Marsch, E., M{\"{u}}hlh{\"{a}}user, K.~H., Schwenn, R., {et~al.}
  1982{\natexlab{b}},
  \href{http://dx.doi.org/10.1029/JA087iA01p00052}{\JournalTitle{Journal of
  Geophysical Research}, 87, 52}

\bibitem[{Marsch \& Tu(2001)}]{Marsch2001}
Marsch, E., \& Tu, C.~Y. 2001,
  \href{http://dx.doi.org/10.1029/2000JA000042}{\JournalTitle{Journal of
  Geophysical Research: Space Physics}, 106, 227}

\bibitem[{Marsch {et~al.}(2003)Marsch, Vocks, \& Tu}]{Marsch2003Onwind}
Marsch, E., Vocks, C., \& Tu, C.~Y. 2003, \JournalTitle{Nonlinear Processes in
  Geophysics}, 10, 101

\bibitem[{Maruca \& Kasper(2013)}]{Maruca2013}
Maruca, B.~A., \& Kasper, J.~C. 2013,
  \href{http://dx.doi.org/10.1016/j.asr.2013.04.006}{\JournalTitle{Advances in
  Space Research}, 52, 723}

\bibitem[{Maruca {et~al.}(2012)Maruca, Kasper, \& Gary}]{Maruca2012}
Maruca, B.~A., Kasper, J.~C., \& Gary, S.~P. 2012,
  \href{http://dx.doi.org/10.1088/0004-637X/748/2/137}{\JournalTitle{The
  Astrophysical Journal}, 748, 137}

\bibitem[{Matteini {et~al.}(2007)Matteini, Landi, Hellinger, Pantellini,
  Maksimovic, Velli, Goldstein, \& Marsch}]{Matteini2007}
Matteini, L., Landi, S., Hellinger, P., {et~al.} 2007,
  \href{http://dx.doi.org/10.1029/2007GL030920}{\JournalTitle{Geophysical
  Research Letters}, 34, L20105}

\bibitem[{Matthaeus \& Goldstein(1982{\natexlab{a}})}]{Matthaeus1982}
Matthaeus, W.~H., \& Goldstein, M.~L. 1982{\natexlab{a}},
  \href{http://dx.doi.org/10.1029/JA087iA08p06011}{\JournalTitle{Journal of
  Geophysical Research}, 87, 6011}

\bibitem[{Matthaeus \& Goldstein(1982{\natexlab{b}})}]{Matthaeus1982b}
---. 1982{\natexlab{b}},
  \href{http://dx.doi.org/doi:10.1029/JA087iA12p10347}{\JournalTitle{Journal of
  Geophysical Research}, 87, 10,347}

\bibitem[{Matthaeus {et~al.}(1982)Matthaeus, Goldstein, \&
  Smith}]{Matthaeus1982a}
Matthaeus, W.~H., Goldstein, M.~L., \& Smith, C.~W. 1982,
  \href{http://dx.doi.org/10.1103/PhysRevLett.48.1256}{\JournalTitle{Physical
  Review Letters}, 48, 1256}

\bibitem[{Moffat(1978)}]{Moffat1978}
Moffat, H.~K. 1978, {Magnetic Field Generation in Electrically Conducting
  Fluids} (Cambridge University Press)

\bibitem[{Montgomery \& Turner(1981)}]{Montgomery1981}
Montgomery, M.~D., \& Turner, L. 1981,
  \href{http://dx.doi.org/10.1063/1.863455}{\JournalTitle{Physics of Fluids},
  24, 825}

\bibitem[{Ogilvie {et~al.}(1995)Ogilvie, Chornay, Fritzenreiter, Hunsaker,
  Keller, Lobell, Miller, Scudder, Sittler, Torbert, Bodet, Needell, Lazarus,
  Steinberg, Tappan, Mavretic, \& Gergin}]{Ogilvie1995}
Ogilvie, K.~W., Chornay, D.~J., Fritzenreiter, R.~J., {et~al.} 1995,
  \href{http://dx.doi.org/10.1007/BF00751326}{\JournalTitle{Space Science
  Reviews}, 71, 55}

\bibitem[{Osman {et~al.}(2014)Osman, Matthaeus, Gosling, Greco, Servidio, Hnat,
  Chapman, \& Phan}]{Osman2014MagneticWind}
Osman, K.~T., Matthaeus, W.~H., Gosling, J.~T., {et~al.} 2014,
  \href{http://dx.doi.org/10.1103/PhysRevLett.112.215002}{\JournalTitle{Physical
  Review Letters}, 112, 1}

\bibitem[{Osman {et~al.}(2013)Osman, Matthaeus, Kiyani, Hnat, \&
  Chapman}]{Osman2013}
Osman, K.~T., Matthaeus, W.~H., Kiyani, K.~H., Hnat, B., \& Chapman, S.~C.
  2013,
  \href{http://dx.doi.org/10.1103/PhysRevLett.111.201101}{\JournalTitle{Physical
  Review Letters}, 111, 1}

\bibitem[{Perri \& Balogh(2010)}]{Perri2010a}
Perri, S., \& Balogh, A. 2010,
  \href{http://dx.doi.org/10.1088/0004-637X/714/1/937}{\JournalTitle{The
  Astrophysical Journal}, 714, 937}

\bibitem[{Perri {et~al.}(2010)Perri, Carbone, \& Veltri}]{Perri2010}
Perri, S., Carbone, V., \& Veltri, P. 2010,
  \href{http://dx.doi.org/10.1088/2041-8205/725/1/L52}{\JournalTitle{The
  Astrophysical Journal Letters}, 725, L52}

\bibitem[{Perri {et~al.}(2012)Perri, Goldstein, Dorelli, \&
  Sahraoui}]{Perri2012DetectionTurbulence}
Perri, S., Goldstein, M.~L., Dorelli, J.~C., \& Sahraoui, F. 2012,
  \href{http://dx.doi.org/10.1103/PhysRevLett.109.191101}{\JournalTitle{Physical
  Review Letters}, 109, 1}

\bibitem[{Podesta(2009)}]{Podesta2009}
Podesta, J.~J. 2009,
  \href{http://dx.doi.org/10.1088/0004-637X/698/2/986}{\JournalTitle{The
  Astrophysical Journal}, 698, 986}

\bibitem[{Podesta \& Gary(2011{\natexlab{a}})}]{Podesta2011a}
Podesta, J.~J., \& Gary, S.~P. 2011{\natexlab{a}},
  \href{http://dx.doi.org/10.1088/0004-637X/742/1/41}{\JournalTitle{The
  Astrophysical Journal}, 742, 41}

\bibitem[{Podesta \& Gary(2011{\natexlab{b}})}]{Podesta2011}
---. 2011{\natexlab{b}},
  \href{http://dx.doi.org/10.1088/0004-637X/734/1/15}{\JournalTitle{The
  Astrophysical Journal}, 734, 15}

\bibitem[{Richardson {et~al.}(1995)Richardson, Paularena, Lazarus, \&
  Belcher}]{Richardson1995}
Richardson, J.~D., Paularena, K.~I., Lazarus, A.~J., \& Belcher, J.~W. 1995,
  \href{http://dx.doi.org/10.1029/94GL03273}{\JournalTitle{Geophysical Research
  Letters}, 22, 325}

\bibitem[{Roberts \& Li(2015)}]{Roberts2015}
Roberts, O.~W., \& Li, X. 2015,
  \href{http://dx.doi.org/10.1088/0004-637X/802/1/1}{\JournalTitle{The
  Astrophysical Journal}, 802, 1}

\bibitem[{Roberts {et~al.}(2015)Roberts, Li, \& Jeska}]{Roberts2015a}
Roberts, O.~W., Li, X., \& Jeska, L. 2015,
  \href{http://dx.doi.org/10.1088/0004-637X/802/1/2}{\JournalTitle{The
  Astrophysical Journal}, 802, 2}

\bibitem[{Roberts {et~al.}(2017)Roberts, Narita, \&
  Escoubet}]{Roberts2017DirectTurbulence}
Roberts, O.~W., Narita, Y., \& Escoubet, C.~P. 2017,
  \href{http://dx.doi.org/10.3847/2041-8213/aa9bf3}{\JournalTitle{The
  Astrophysical Journal}, 851}

\bibitem[{Russell(1972)}]{Russell1972}
Russell, C.~T. 1972, in NASA Special Pub. 308, Solar Wind, 365

\bibitem[{Sahraoui {et~al.}(2010)Sahraoui, Goldstein, Belmont, Canu, \&
  Rezeau}]{Sahraoui2010}
Sahraoui, F., Goldstein, M.~L., Belmont, G., Canu, P., \& Rezeau, L. 2010,
  \href{http://dx.doi.org/10.1103/PhysRevLett.105.131101}{\JournalTitle{Physical
  Review Letters}, 105}

\bibitem[{Schekochihin {et~al.}(2009)Schekochihin, Cowley, Dorland, Hammett,
  Howes, Quataert, \& Tatsuno}]{Schekochihin2009}
Schekochihin, A.~A., Cowley, S.~C., Dorland, W., {et~al.} 2009,
  \href{http://dx.doi.org/10.1088/0067-0049/182/1/310}{\JournalTitle{The
  Astrophysical Journal Supplement Series}, 182, 310}

\bibitem[{Schwenn(1990)}]{Schwenn1990}
Schwenn, R. 1990, in Physics of the Inner Heliosphere I (Springer Physics and
  Chemistry in Space), 99

\bibitem[{Servidio {et~al.}(2014)Servidio, Osman, Valentini, Perrone, Califano,
  Chapman, Matthaeus, \& Veltri}]{Servidio2014}
Servidio, S., Osman, K.~T., Valentini, F., {et~al.} 2014,
  \href{http://dx.doi.org/10.1088/2041-8205/781/2/L27}{\JournalTitle{The
  Astrophysical Journal}, 781, L27}

\bibitem[{Smith(2003)}]{Smith2003MagneticWind}
Smith, C.~W. 2003,
  \href{http://dx.doi.org/10.1016/S0273-1177(03)00770-1}{\JournalTitle{Advances
  in Space Research}, 32, 1971}

\bibitem[{Smith {et~al.}(2006)Smith, Hamilton, Vasquez, \& Leamon}]{Smith2006}
Smith, C.~W., Hamilton, K., Vasquez, B.~J., \& Leamon, R.~J. 2006,
  \href{http://dx.doi.org/10.1086/506151}{\JournalTitle{The Astrophysical
  Journal}, 645, L85}

\bibitem[{Smith {et~al.}(2001)Smith, Matthaeus, Zank, Ness, Oughton, \&
  Richardson}]{Smith2001}
Smith, C.~W., Matthaeus, W.~H., Zank, G.~P., {et~al.} 2001,
  \href{http://dx.doi.org/10.1029/2000JA000366}{\JournalTitle{Journal of
  Geophysical Research}, 106, 8253}

\bibitem[{Smith {et~al.}(2012)Smith, Vasquez, \& Hollweg}]{Smith2012}
Smith, C.~W., Vasquez, B.~J., \& Hollweg, J.~V. 2012,
  \href{http://dx.doi.org/10.1088/0004-637X/745/1/8}{\JournalTitle{The
  Astrophysical Journal}, 745, 8}

\bibitem[{Stawarz {et~al.}(2009)Stawarz, Smith, Vasquez, Forman, \&
  MacBride}]{Stawarz2009}
Stawarz, J.~E., Smith, C.~W., Vasquez, B.~J., Forman, M.~A., \& MacBride, B.~T.
  2009, \href{http://dx.doi.org/10.1088/0004-637X/697/2/1119}{\JournalTitle{The
  Astrophysical Journal}, 697, 1119}

\bibitem[{Stix(1992)}]{Stix1992}
Stix, T.~H. 1992, {Waves in Plasmas} (American Institute of Physics)

\bibitem[{Taylor(1938)}]{Taylor1938}
Taylor, G.~I. 1938,
  \href{http://dx.doi.org/10.1098/rspa.1938.0032}{\JournalTitle{Proceedings of
  the Royal Society A: Mathematical and Physical Sciences}, 164, 476}

\bibitem[{Telloni \& Bruno(2016)}]{Telloni2016}
Telloni, D., \& Bruno, R. 2016,
  \href{http://dx.doi.org/10.1093/mnrasl/slw135}{\JournalTitle{Monthly Notices
  of the Royal Astronomical Society: Letters}, 463, L79}

\bibitem[{Telloni {et~al.}(2012)Telloni, Bruno, D'Amicis, Pietropaolo, \&
  Carbone}]{Telloni2012}
Telloni, D., Bruno, R., D'Amicis, R., Pietropaolo, E., \& Carbone, V. 2012,
  \href{http://dx.doi.org/10.1088/0004-637X/751/1/19}{\JournalTitle{The
  Astrophysical Journal}, 751, 19}

\bibitem[{Telloni {et~al.}(2015)Telloni, Bruno, \& Trenchi}]{Telloni2015}
Telloni, D., Bruno, R., \& Trenchi, L. 2015,
  \href{http://dx.doi.org/10.1088/0004-637X/805/1/46}{\JournalTitle{The
  Astrophysical Journal}, 805, 46}

\bibitem[{Telloni {et~al.}(2013)Telloni, Perri, Bruno, Carbone, \&
  D'Amicis}]{Telloni2013}
Telloni, D., Perri, S., Bruno, R., Carbone, V., \& D'Amicis, R. 2013,
  \href{http://dx.doi.org/10.1088/0004-637X/776/1/3}{\JournalTitle{The
  Astrophysical Journal}, 776, 3}

\bibitem[{TenBarge \& Howes(2012)}]{TenBarge2012EvidenceSimulations}
TenBarge, J.~M., \& Howes, G.~G. 2012,
  \href{http://dx.doi.org/doi:http://dx.doi.org/10.1063/1.3693974}{\JournalTitle{Physics
  of Plasmas}, 19, 55901}

\bibitem[{TenBarge {et~al.}(2012)TenBarge, Podesta, Klein, \&
  Howes}]{TenBarge2012InterpretingWind}
TenBarge, J.~M., Podesta, J.~J., Klein, K.~G., \& Howes, G.~G. 2012,
  \href{http://dx.doi.org/10.1088/0004-637X/753/2/107}{\JournalTitle{The
  Astrophysical Journal}, 753, 107}

\bibitem[{Torrence \& Compo(1998)}]{Torrence1998}
Torrence, C., \& Compo, G.~P. 1998,
  \href{http://dx.doi.org/10.1175/1520-0477(1998)079<0061:APGTWA>2.0.CO;2}{\JournalTitle{Bulletin
  of the American Meteorological Society}, 79, 61}

\bibitem[{Tsurutani {et~al.}(1994)Tsurutani, Arballo, Mok, Smith, Mason, \&
  Tan}]{Tsurutani1994ElectromagneticObservations}
Tsurutani, B.~T., Arballo, J.~K., Mok, J., {et~al.} 1994,
  \href{http://dx.doi.org/10.1029/94GL00566}{\JournalTitle{Geophysical Research
  Letters}, 21, 633}

\bibitem[{Tu \& Marsch(1995)}]{Tu1995}
Tu, C.~Y., \& Marsch, E. 1995,
  \href{http://dx.doi.org/10.1007/BF00748891}{\JournalTitle{Space Science
  Reviews}, 73, 1}

\bibitem[{Tu \& Marsch(2001)}]{Tu2001OnCorona}
---. 2001, \href{http://dx.doi.org/10.1029/2000JA000024}{\JournalTitle{Journal
  of Geophysical Research}, 106, 8233}

\bibitem[{Tu \& Marsch(2002)}]{Tu2002}
---. 2002, \href{http://dx.doi.org/10.1029/2001JA000150}{\JournalTitle{Journal
  of Geophysical Research: Space Physics}, 107, 1}

\bibitem[{Vasquez \& Markovskii(2012)}]{Vasquez2012VelocityRegime}
Vasquez, B.~J., \& Markovskii, S.~A. 2012,
  \href{http://dx.doi.org/10.1088/0004-637X/747/1/19}{\JournalTitle{The
  Astrophysical Journal}, 747, 19}

\bibitem[{Wang {et~al.}(2018)Wang, Tu, He, \&
  Wang}]{Wang2018Ion-scaleTurbulence}
Wang, X., Tu, C.~Y., He, J., \& Wang, L.-H. 2018,
  \href{http://dx.doi.org/10.1002/2017JA024813}{\JournalTitle{Journal of
  Geophysical Research: Space Physics}, 123}

\bibitem[{Wicks {et~al.}(2012)Wicks, Forman, Horbury, \& Oughton}]{Wicks2012}
Wicks, R.~T., Forman, M.~A., Horbury, T.~S., \& Oughton, S. 2012,
  \href{http://dx.doi.org/10.1088/0004-637X/746/1/103}{\JournalTitle{The
  Astrophysical Journal}, 746, 103}

\bibitem[{Wicks {et~al.}(2010)Wicks, Horbury, Chen, \&
  Schekochihin}]{Wicks2010}
Wicks, R.~T., Horbury, T.~S., Chen, C. H.~K., \& Schekochihin, A.~A. 2010,
  \href{http://dx.doi.org/10.1111/j.1745-3933.2010.00898.x}{\JournalTitle{Monthly
  Notices of the Royal Astronomical Society: Letters}, 407, L31}

\bibitem[{Wicks {et~al.}(2016)Wicks, Alexander, Stevens, Wilson~III, Moya,
  Vi{\~{n}}as, Jian, Roberts, O’Modhrain, Gilbert, \& Zurbuchen}]{Wicks2016}
Wicks, R.~T., Alexander, R.~L., Stevens, M.~L., {et~al.} 2016,
  \href{http://dx.doi.org/10.3847/0004-637X/819/1/6}{\JournalTitle{The
  Astrophysical Journal}, 819, 6}

\bibitem[{Woltjer(1958)}]{Woltjer1958a}
Woltjer, L. 1958,
  \href{http://www.pnas.org/content/44/6/489}{\JournalTitle{Proceedings of the
  National Academy of Sciences}, 44, 489}

\end{thebibliography}

\appendix

\section{MFI Noise Floor Determination} \label{sec:appB}

The trajectory of the \textit{Wind} spacecraft allows us to determine the MFI instrument noise-floor using \textit{in-situ} data. To measure the noise level at high frequencies, we require measurements of `quiet', smoothly-varying magnetic field so that there are as few physical fluctuations as possible at these frequencies so that noise completely dominates the measured signal. The solar wind is largely unsuitable for this due to the presence of broadband turbulent fluctuations. However, the spacecraft spent considerable time in the Earth's magnetosphere before 2005. In particular, during early 2004, WIND made several passes through the tail-lobes of the far magnetotail. These high-latitude regions surrounding the central plasma sheet have characteristic low plasma density and stretched-out field lines, and so there are few high-frequency fluctuations in the field.

\begin{figure}[h]
\centering
\includegraphics[scale=0.35]{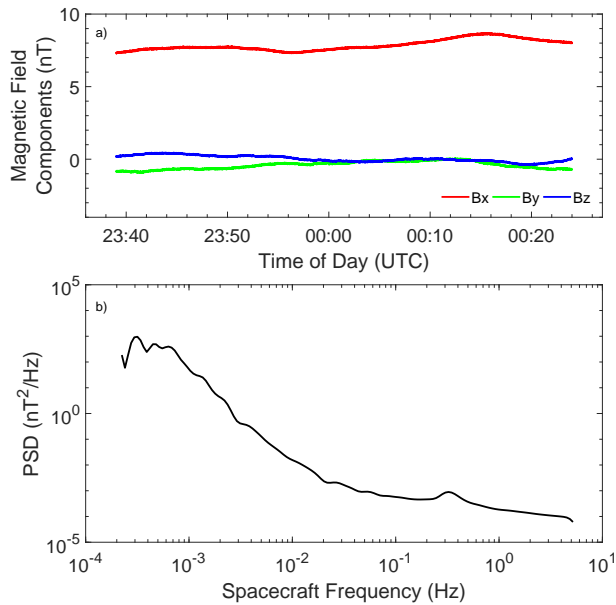}
\caption{(a) Magnetic field components in GSE coordinates from one of our selected intervals for determining the noise-floor (23:39-00:24 12-13th February 2004). (b) The corresponding PSD of the trace power spectrum of the magnetic field for the same interval, highlighting the flattening of the spectrum at high frequencies due to noise, as well as the peak at about 0.33 Hz due to spacecraft spin.}
\label{fig:A}
\end{figure}

The MFI instrument has several dynamic range-gates which are used to measure the magnetic field vector components within a specific range of values; the default is $\pm$16 nT \citep{Lepping1995}. When one of the field components exceeds an amplitude of 14 nT, there is a transition to the next gate, $\pm$64 nT. It is likely that the noise level at high frequencies has a different amplitude for each range-gate since the digital resolution decreases by a factor of 4 for each consecutive range-gate. This effect should increase the noise level by a factor of about 2 for each gate. Despite this, we consider the noise-floor only for the default range-gate, $\pm$16 nT, since the magnetic field components typically do not exceed $\pm$14 nT during quiet conditions in the solar wind. In the rare instance that the magnetic field exceeds this threshold, we use a conservative SNR of 10 that will minimize any impact on our results.

Using the criteria we have discussed, we identify 89 suitable tail-lobe intervals to determine the noise-floor. We show the magnetic field time series from an example interval from early February 2004 in Figure 10(a), and its corresponding power spectrum in Figure 10(b), revealing the noise floor at high frequencies and illustrating the quietness of our selected period. To find the instrument noise level, we calculate the PSD using Equation (\ref{equ:psd}) and average over each interval, using the method documented in Section \ref{sec:spectra}, padding each interval to remove any border effects. All identified periods are between 30 minutes and several hours, long enough to provide a stable estimate of the PSD. We refine the selected intervals down to a final 22 periods by removing datasets with signals attributed to physical fluctuations above 0.1 Hz and average the PSD estimates for all intervals together to give one final estimate for the noise-floor. We show the PSDs of all 22 periods in Figure 11, where the black line shows the average and the red line the original noise-floor estimate published by \citet{Lepping1995}, performed on a prototype sensor before launch.

\begin{figure}[h]
\centering
\includegraphics[scale=0.4]{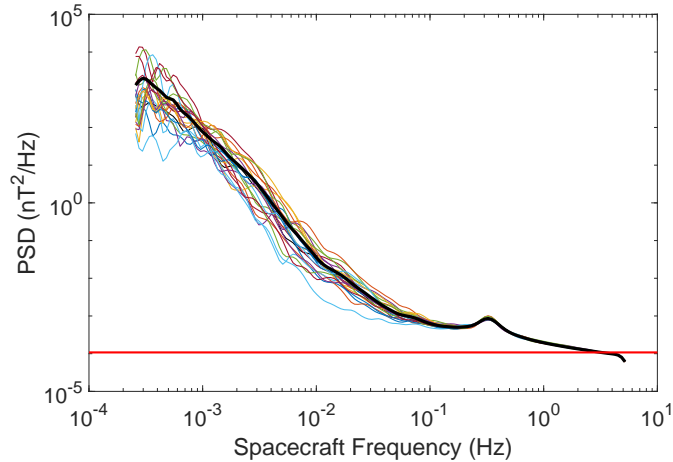}
\caption{The PSD of the final 22 datasets used for the combined average MFI noise-floor estimate, given by the black line. The red line is the original determination by \citet{Lepping1995}.}
\label{fig:B}
\centering
\end{figure}

The major difference in our noise-floor spectrum compared with the original estimate is the peak at 0.33 Hz. We attribute this peak to the spin of the spacecraft every 3 seconds. This peak is notch filtered as described by \citet{Koval2013} to remove most of this artifact, but some residual power remains. At higher frequencies to the peak, we find a power law fit of $\textnormal{PSD}=a f\,^b$ where $a=1.944\times10^{-4}$ and $b=-0.5328$ before the amplitude eventually coincides with the original estimate at about 3 Hz. This high-frequency part of the spectrum is due to the aliasing of the spin tone harmonics \citep{Koval2013}, which cannot be removed by filtering, as well as noise from the digitization process \citep{Bennett1948SpectraSignals,Russell1972}. Another source of noise is the aliasing of power due to the presence of turbulent fluctuations measured at higher frequencies than the Nyquist frequency (5.4 Hz) of the MFI instrument \citep{Russell1972,Klein2014}. This effect cannot be completely removed by our noise-floor treatment here, however, our estimate works well with solar wind data. Approaching the Nyquist frequency, the spectrum steepens again, but this is due to effects of the wavelet transform. We take the amplitude of power at frequencies $<$0.1 Hz as due to physical fluctuations in the magnetic field and not instrumental- or spacecraft-induced noise.

Although not shown here, we also identified intervals from 2000-2003 to calculate the noise-floor, but these show no substantial variation in the noise level compared to the results from 2004. The differences between our noise-floor estimate and the original estimate by \citet{Lepping1995} show that the noise level between 0.1-1 Hz is greater than initially thought, highlighting its importance when investigating turbulent phenomena in the solar wind at these frequencies. We use the amplitude of the PSD between 0.1-5 Hz as the noise-floor in our analysis of the turbulent magnetic field fluctuations in the solar wind, incorporating both the peak due to spacecraft spin and the high-frequency power-law component. We provide our dataset of the noise-floor at these frequencies as supplementary material to this paper for use in future studies.

\end{document}